\documentclass[conference]{IEEEtran}
\IEEEoverridecommandlockouts
\usepackage{tikz}
\usepackage{amsmath}
\usepackage{soul}
\usepackage{multirow}
\usepackage{adjustbox}
\usepackage{enumitem}
\setlist[itemize]{itemsep=0pt}
\usepackage{amssymb}

\usepackage{hyperref}
\usepackage{algorithm}
\usepackage{algpseudocode}
\usepackage{adjustbox}
\usepackage{pgfplots}
\usepackage{filecontents}
\usepackage{subcaption}
\pgfplotsset{compat=1.18}
\usepackage{adjustbox}
\usepackage{xurl}
\usepackage{mathastext}

\def\BibTeX{{\rm B\kern-.05em{\sc i\kern-.025em b}\kern-.08em
    T\kern-.1667em\lower.7ex\hbox{E}\kern-.125emX}}
\begin{document}

\title{ Intra-Section Code Cave Injection for Adversarial \\ Evasion Attacks on Windows PE Malware File}

\author{\IEEEauthorblockN{1\textsuperscript{st} Kshitiz Aryal}
\IEEEauthorblockA{\textit{Department of Computer Science} \\
\textit{Tennessee Tech University}\\
Cookeville, USA \\
karyal42@tntech.edu}
\and
\IEEEauthorblockN{2\textsuperscript{nd} Maanak Gupta}
\IEEEauthorblockA{\textit{Department of Computer Science} \\
\textit{Tennessee Tech University}\\
Cookeville, USA \\
mgupta@tntech.edu }
\and
\IEEEauthorblockN{3\textsuperscript{rd} Mahmoud Abdelsalam}
\IEEEauthorblockA{\textit{Department of Computer Science} \\
\textit{North Carolina A\&T State University}\\
Greensboro, NC \\
mabdelsalam1@ncat.edu}
\and
\IEEEauthorblockN{4\textsuperscript{th} Moustafa Saleh}
\IEEEauthorblockA{\textit{Oracle Cloud Infrastructure} \\
\textit{Oracle}\\
Seattle, USA \\
moustafa.saleh@oracle.com}
}

\maketitle

\begin{abstract}
Windows malware is predominantly available in cyberspace and is a prime target for deliberate adversarial evasion attacks. Although researchers have investigated the adversarial malware attack problem, a multitude of important questions remain unanswered, including (a) Are the existing techniques to inject adversarial perturbations in Windows Portable Executable (PE) malware files effective enough for evasion purposes?; (b) Does the attack process preserve the original behavior of malware?; (c) Are there unexplored approaches/locations that can be used to carry out adversarial evasion attacks on Windows PE malware?; and (d) What are the optimal locations and sizes of adversarial perturbations required to evade an ML-based malware detector without significant structural change in the PE file? To answer some of these questions, this work proposes a novel approach that injects a code cave within the section (i.e., intra-section) of Windows PE malware files to make space for adversarial perturbations. In addition, a code loader is also injected inside the PE file, which reverts adversarial malware to its original form during the execution, preserving the malware's functionality and executability. To understand the effectiveness of our approach, we injected adversarial perturbations inside the \texttt{.text}, \texttt{.data} and \texttt{.rdata} sections, generated using the gradient descent and Fast Gradient Sign Method (FGSM) to target the two popular CNN-based malware detectors, MalConv and MalConv2. Our experimental analysis yielded impressive results, achieving an evasion rate of 92.31\% with gradient descent and 96.26\% with FGSM when targeting MalConv, as compared to the evasion rate of 16.17\% for append attacks. Similarly, in the case of an attack against MalConv2, our approach achieved a remarkable maximum evasion rate of 97.93\% with gradient descent and 94.34\% with FGSM, significantly surpassing the 4.01\% and 54.75\% evasion rate observed with append attacks. The work concludes by highlighting the challenges of current state-of-the-art approaches in crafting adversarial malware and proposing future research directions.
\end{abstract}


\section{Introduction}
\label{sec:Introduction}

\begin{table*}[!t]
\centering
\caption{Comparison of existing adversarial evasion attacks on Windows malware}
\vspace{-3mm}
\label{Table:adv_evasion_list}
\begin{adjustbox}{width=\textwidth}
\begin{tabular}{|c|c|c|c|c|c|c|c|c|c|l|l|l|c|c|}
\hline
\multirow{2}{*}{\textbf{Research Works}} & \multirow{2}{*}{\textbf{\begin{tabular}[c]{@{}c@{}}Target\\ Model\end{tabular}}} & \multicolumn{2}{c|}{\textbf{\begin{tabular}[c]{@{}c@{}}Attacker's \\ Knowledge\end{tabular}}} & \multicolumn{2}{c|}{\textbf{\begin{tabular}[c]{@{}c@{}}Perturbation\\ Space\end{tabular}}} &  \multicolumn{4}{c|}{\textbf{\begin{tabular}[c]{@{}c@{}}Attack Location\\ in Windows PE\end{tabular}}}  & \multicolumn{2}{c|}{\textbf{\begin{tabular}[c]{@{}c@{}}Code\\ Caves\end{tabular}}}                  & \multirow{2}{*}{\rotatebox[origin=c]{90}{\textbf{Flexibility}}} & \multirow{2}{*}{\rotatebox[origin=c]{90}{\textbf{\begin{tabular}[c]{@{}c@{}}Code Loader\\ Injection\end{tabular}}}} & \multirow{2}{*}{\rotatebox[origin=c]{90}{\textbf{\begin{tabular}[c]{@{}c@{}}Preserves\\ Functionality\end{tabular}}}} \\ \cline{3-12}
                                  &                                  & \multicolumn{1}{p{6.5mm}|}{\rotatebox[origin=c]{90}{\textbf{White Box}}}                  & \rotatebox[origin=c]{90}{\textbf{Black Box}}                 & \multicolumn{1}{p{7mm}|}{\rotatebox[origin=c]{90}{\textbf{Problem Space}}}         & \rotatebox[origin=c]{90}{\textbf{Feature Space}}           & \multicolumn{1}{l|}{\rotatebox[origin=c]{90}{\textbf{End of File}}} & \multicolumn{1}{l|}{\rotatebox[origin=c]{90}{\textbf{Headers}}} & \multicolumn{1}{l|}{\rotatebox[origin=c]{90}{\textbf{Inter-Sections}}} & \rotatebox[origin=c]{90}{\textbf{Intra-Section}} & \multicolumn{1}{l|}{\rotatebox[origin=c]{90}{\textbf{Existing}}} & \rotatebox[origin=c]{90}{\textbf{Injecting}} &                                       &                                     &                                                                                             \\ \hline
Anderson et al.~\cite{anderson2018learning} & GBDT & & $\surd$ & $\surd$ & & $\surd$ & & & & $\surd$ & & & & $\surd$ \\ \hline
Hu et al.~\cite{hu2017black} & RNN & & $\surd$ & &$\surd$ & &&&&&&&&\\ \hline
Rosenberg et al.~\cite{rosenberg2018generic} & RNN & & $\surd$ & &$\surd$ &&&&&&&&&\\ \hline
Kreuk et al.~\cite{kreuk2018adversarial} & CNN (MalConv) & $\surd$ &&$\surd$&&$\surd$&&&&&&&&$\surd$ \\ \hline
Kolosnjaji et al. ~\cite{kolosnjaji2018adversarial} & CNN (MalConv) & $\surd$ & & $\surd$ &&$\surd$&&&&&&&&$\surd$ \\ \hline
Suciu et al. ~\cite{suciu2019exploring} & CNN (MalConv) & $\surd$ &&$\surd$&&$\surd$&&$\surd$&&$\surd$&&&&$\surd$\\ \hline
Khormali et al.~\cite{khormali2019copycat} & CNN & $\surd$ &&$\surd$&&&&&&&&&&\\ \hline
Demetrio et al. ~\cite{demetrio2019explaining} & CNN (MalConv) & $\surd$&&$\surd$&&&$\surd$&&&$\surd$&&&&\\ \hline
Demetrio et al.~\cite{demetrio2021functionality} & GBDT, MalConv&&$\surd$&$\surd$&&$\surd$&$\surd$&&&$\surd$&&&&$\surd$\\ \hline 
Yuste et al.~\cite{yuste2022optimization} & CNN (MalConv) & &$\surd$&$\surd$&&&&$\surd$&&&$\surd$&$\surd$&&$\surd$ \\ \hline
\textbf{Our Approach} &\textbf{ MalConv, MalConv2} & ${\surd}$&&$\surd$&&&&&$\surd$&&$\surd$&$\surd$&$\surd$&$\surd$\\ \hline
\end{tabular}
\end{adjustbox}
\vspace{-4mm}
\end{table*}

A recent vulnerability and a growing threat to the machine learning community has been exposed in the form of \textit{adversarial attacks}. A small well-directed synthetic perturbation can easily \textit{fool} a sophisticated deep learning classifier, revealing its weakness against attacks of similar nature ~\cite{biggio2013evasion, goodfellow2014explaining} in different application domains.
When added to an image, these small perturbations do not change the perception of the human oracle but make a significant shift in the classifier's feature space, eventually disguising the classifier model. These small perturbations can be introduced in the training data as an adversarial \textit{poisoning} attack or in testing data as an adversarial \textit{evasion} attack. However, in practical scenarios, attackers rarely have access to the training data of the ML model; therefore, adversarial evasion attacks done by modifying the test sample are more realistic and easier for an attacker, and hence is the focus of this work. 

To formally define an Adversarial Evasion (AE) attack, consider a machine learning classifier $C$ with test sample $x$ such that $C\left(x\right) = y_{real}$, where $y_{real}$ is the true label of $x$. Now, the adversarial perturbation $\delta$ is introduced to a test sample $x$ in such a way that $C\left(x \texttt{+} \delta\right) = y_{adv}$, where $y_{adv}$ is the new adversarial label of $x$ and $y_{real} \neq y_{adv}$. In this definition, the role of an attacker is to introduce small imperceptible perturbation $\delta$ sufficient to change the label of $x$ from  $y_{real}$ to $y_{adv}$. 
These adversarial evasion attacks, initially originating from the image domain, have propagated to several fields, including speech recognition~\cite{alzantot2018did, qin2019imperceptible}, signal processing~\cite{sadeghi2018adversarial, kim2021channel}, medical analysis~\cite{finlayson2019adversarial, ma2021understanding}, malware detection~\cite{anderson2018learning, kreuk2018adversarial,aryal2022analysis}, etc. Regardless of the difference in the domain, the goal of the AE attack remains the same: fooling a trained ML-based classifier using a modified adversarial test sample.

In malware, these adversarial evasion attacks are still far from facing the real challenges of dealing with malware in the wild. The crafted perturbations must be inserted into the malware file such that a trained ML-based detector confuses the malware file as \textit{benign}. However, the strict semantic constraints of a binary executable make adversarial evasion attacks a completely different and complex game in malware as compared to other domains. While carrying out these attacks on malware, an attacker could unintentionally introduce perturbations to the malware file that can easily cease the execution and functionality of malware, rendering it useless. This significant constraint has restricted the research in fully exploiting a malware file with advanced adversarial attacks. Based on our literature review and analysis, the current progress in adversarial evasion attacks for malware binaries can be broadly categorized into two: (1) All the possibilities of creating evasion attacks have not been fully exploited; (2) Presence of significant research gaps in malware-specific adversarial attacks in a practical, real-world environment.

\textbf{Motivation and Existing Limitations}: Windows malware is the most common threat found in cyberspace~\cite{95ofalln3:online}, rendering it an appealing attack vector used by adversaries. 
Many initial AE attacks were carried out in \textit{feature space}~\cite{hu2017black,rosenberg2018generic} and were limited due to the inability to map features back to \textit{problem space} (i.e., functional malware). The problem space refers to the PE file where it exists in its standard defined format, whereas the feature space can be static or dynamic features obtained from some analysis. Since it is difficult to accurately map the feature space to the problem space in malware, it becomes nearly impossible to generate functional malware from the adversarial perturbations generated in the feature space.
As such, recent research shifted to working directly on the malware PE file (i.e., problem space). Several existing research (as shown in Table \ref{Table:adv_evasion_list}) have focused on the locations inside the Windows PE files where adversarial perturbations can be inserted~\cite{suciu2019exploring, demetrio2019explaining, yuste2022optimization}. These locations are critical and significantly impact preserving the functionality, hiding the perturbations, and the performance of malware detectors~\cite{aryal2021survey}. However, current malware AE attack approaches lack \textit{flexibility}, as most of the attacks are carried out either at the end of the file~\cite{anderson2018learning, kreuk2018adversarial, kolosnjaji2018adversarial} or at the existing empty spaces (i.e., code caves) inside the file~\cite{suciu2019exploring, demetrio2019explaining}. This limits the attacker's ability to inject perturbations freely across the different regions of a malware file. 
Another critical aspect in carrying out an AE attack is that preserving the \textit{functionality} of a malware file is either overlooked or insufficiently investigated. In simpler words, the current attacks are either prone to breaking the malware file~\cite{khormali2019copycat, demetrio2019explaining} or are too cautious and simplistic in modifying the malware by adding perturbations only at the end of the file. To the best of our knowledge, only one approach proposed by Yuste et al.~\cite{yuste2022optimization} provided more flexibility in terms of the amount of empty space for perturbation. However, their approach relied on creating empty regions between the sections of binary malware files, leading to the insertion of adversarial perturbation in pronounced regions that can easily be red-flagged by the malware detectors. Further, Yuste et al.'s~\cite{yuste2022optimization} work uses a genetic algorithm to generate adversarial perturbations, which is a blackbox approach and computationally very expensive. Genetic algorithms are knowledge agnostic and do not always use the available knowledge of the problem domain to get to the solution.


\textbf{Proposed Approach:} 
In this work, our goal is to create an adversarial evasion attack by inserting \textit{intra-section} code cave and perturbation in a Windows PE file, targeting a widely used Convolutional Neural Network (CNN) based, MalConv~\cite{raff2018malware} and MalConv2~\cite{raff2021classifying} models, that has been an academic standard for performing AE attacks~\cite{kolosnjaji2018adversarial, suciu2019exploring, kreuk2018adversarial}.
This work addresses the limited flexibility of existing work~\cite{anderson2018learning, kreuk2018adversarial, kolosnjaji2018adversarial} for injecting the adversarial perturbation by dynamically introducing \textit{intra-section} code caves. These code caves are empty spaces inside PE file sections where the adversarial perturbations of the desired size can be injected.
Further, as these code caves are introduced inside the malware sections, it also helps to hide the adversarial perturbations from the obvious regions easily flagged by the malware detector as in ~\cite{yuste2022optimization}. In addition, our approach also resolves the challenge of preserving the \textit{functionality} and \textit{executability} of malware during modification ~\cite{suciu2019exploring, demetrio2019explaining} by injecting a code loader that restores the original malware during execution. The code loader is capable of erasing a code cave and an adversarial perturbation when malware executes, thus preserving the behavior of the malware. Also, since our approach carries out AE attacks in problem space, the issue of mapping between problem space and feature space ~\cite{hu2017black, rosenberg2018generic} is nonexistent. Table ~\ref{Table:adv_evasion_list} differentiates our approach from the existing works.

As detailed in our experimental results, we analyzed the effectiveness of \textit{intra-section} adversarial evasion attacks against baseline append attacks on 1000 Windows PE malware files by adding perturbations inside the code caves, dynamically injected within the \texttt{.text}, \texttt{.data}, and \texttt{.rdata} sections. We used gradient descent and FGSM algorithms to generate adversarial perturbation while targeting two popular CNN-based malware detectors MalConv~\cite{raff2018malware} and MalConv2~\cite{raff2021classifying}, extensively used in literature for malware evasion research. These attacks on \texttt{.text}, \texttt{.data}, and \texttt{.rdata} sections produced an evasion rate of 63.45\%, 69.77\%, and 92.31\%, respectively, using gradient descent as compared to the evasion rate of 16.17\% produced by baseline append attacks against the target MalConv model. Similarly, our experiments reached the maximum evasion of 96.26\% with the FGSM algorithm against MalConv. On targeting MalConv2, we achieved a maximum evasion rate of 97.93\% through intra-section code cave insertion using gradient descent as compared to 4.01\% with append attack. Likewise, our approach with FGSM achieved a maximum evasion rate of 94.34\% against MalConv2.  Our \textit{intra-section} approach also resulted in a higher confidence reduction in both MalConv and MalConv2 as compared to append attacks. The main contributions of this work are:
\begin{itemize}[leftmargin = *,noitemsep,topsep=0pt]
    \item We propose a novel adversarial evasion attack that introduces an approach to inject and hide the adversarial perturbations within the section (intra-section) of a Windows PE malware file by introducing a new code cave.
    \item We add a code loader into the PE file that restores the original malware during the execution, hence preserving its functionality and executability.
    \item We test the effectiveness of our proposed approach by attacking different sections (\texttt{.text}, \texttt{.data}, and \texttt{.rdata}) of Windows PE malware file while comparing against baseline append attacks. 
    \item In a comprehensive assessment, we scrutinized our approach using two distinct adversarial generation algorithms, Gradient Descent and the Fast Gradient Sign Method, while simultaneously targeting two widely recognized CNN-based malware detectors, MalConv and MalConv2.
    \item We demonstrate that our intra-section adversarial evasion attack achieved an evasion rate of up to 96\% against MalConv and 97\% against MalConv2 by injecting only 15\% perturbation relative to the malware file size. 
    \item We discuss the challenges and limitations of our approach while highlighting possible future directions.
\end{itemize}

The rest of the paper is organized as below. Section \ref{sec:background} covers the relevant background and related work. The threat model for the AE attacks discussing attack surface, attackers' knowledge, capabilities, and goals are presented in Section \ref{sec:Threat_model}. Detailed methodology of our approach, providing a step-by-step guide to generate an intra-section code cave and inserting perturbation in Windows PE file, is in Section \ref{Sec:Methodology}.  
Section \ref{Sec: Experimental_results} discusses the dataset, experimental setup, and results. We elaborate on the challenges and limitations in Section \ref{sec:Discussion}. Finally, we conclude this work in Section \ref{Sec:Conclusion}.

\section{Background and Related Work}
\label{sec:background}
In this section, we will discuss relevant preliminaries and literature for adversarial malware attacks. We also highlight the limitations of existing works and how our approach is novel and fundamentally different from prior research.

\subsection{Malware Detection}
\label{subsec:Malware_det}

\begin{figure}[!t]
    \centering
    \includegraphics[scale=0.75]{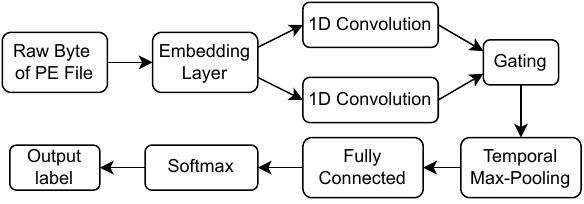}
    \vspace{-2.5mm}
    \caption{Architecture of MalConv \cite{raff2018malware}}
    \label{fig:malconv}
\end{figure}

The ability of machine learning to generalize their knowledge over the features taken from the static or dynamic analysis has edged them way ahead of existing rule-based approaches. Machine learning models have excelled, especially in detecting previously unseen or unknown malware. Machine learning techniques like deep learning and Convolutional Neural Networks (CNN) can automatically learn the features and representations from the raw binary of the executable~\cite{sl2019windows, sharma2019malware, raff2018malware}. The automatically learned features remove the overhead of manual work while also making the malware detector resistant to obfuscation techniques. One of the popular 1-d convolutional neural network-based malware detectors, MalConv ~\cite{raff2018malware}, combines convolutional activation with a global max-pooling layer followed by a fully connected layer, as shown in Figure \ref{fig:malconv}. This architecture helps to produce the activation without depending on the location of its detected features. The embedding layer first encodes the input raw bytes into a fixed-size embedding vector, an input to the convolutional layer. The convolutional layers extract features from the input vector, and the pooling layer shrinks the feature-space dimension. Finally, the fully connected layer is the non-linear classifier for the given feature representation of the input. 

\begin{figure}[!t]
    \centering
    \includegraphics[width=.9\linewidth]{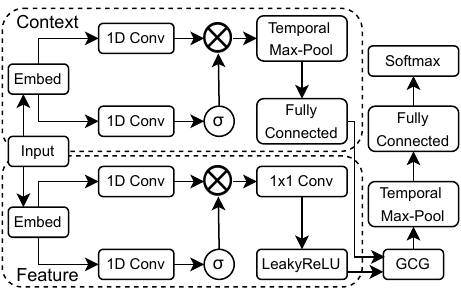}
    \vspace{-3mm}
    \caption{Architecture of MalConv2 \cite{raff2021classifying}}
    \label{fig:malconv2}
\end{figure}

Raff et al. improved original MalConv~\cite{raff2018malware} and proposed MalConv2~\cite{raff2021classifying} whose architecture is shown in Figure \ref{fig:malconv2}. The improvement involves the development of a novel Global Channel Gating mechanism, introducing an attention mechanism that efficiently learns feature interactions across an impressive 100 million time steps- an aspect that was absent in the original MalConv model. 
Unlike MalConv, which was constrained to processing files of up to 2MB, MalConv2~\cite{raff2021classifying} removes the limitation, expanding its capabilities to handle larger files up to 16MB. Our work uses both the MalConv and more robust MalConv2 malware detectors as our target models for AE attacks.


\begin{figure}[!t]
    \centering
    \begin{subfigure}{0.50\linewidth}
        \includegraphics[width=.9\linewidth]{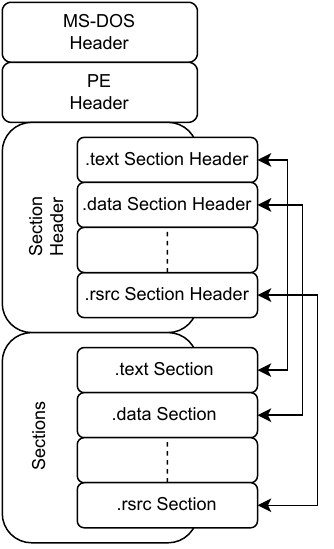} 
         \caption{}
        \label{fig:Windows_PE_Struct}   
    \end{subfigure}
    \hfill
    \begin{subfigure}{0.40\linewidth}
         \includegraphics[width=.9\linewidth]{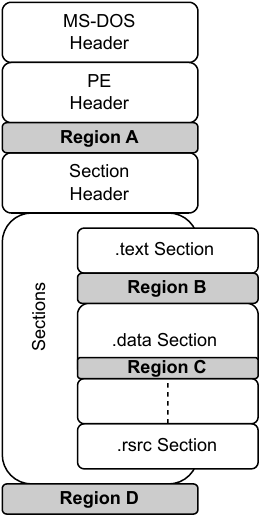}
         \caption{}
         \label{fig:pe_location}
    \end{subfigure}
    \vspace{-2.5mm}
    \caption{(a) Structure of Windows PE file, (b) Different locations inside Windows PE file for injecting perturbation}
    \label{fig:evasioat}
    \vspace{-5mm}
\end{figure}

\subsection{Windows PE File Structure}
\vspace{-2mm}
\label{subsec:Windows_PE}
The Portable Executable (PE) file format gives information for the Windows Operating System (OS) loader to load the file into memory and execute it. The format for PE should not make it architecture-specific~\cite {PEFormat58_online}. The executable format is based on the Common Object File Format (COFF) composed of linear data streams. The components of a Windows PE file, as shown in Figure \ref{fig:Windows_PE_Struct} consist of different information about the file. The header sections consist of metadata about the file, whereas the main body consists of codes, data, and resources. The file starts with an MS-DOS header, a real-mode program stub, a PE File signature, and an optional header. These headers include information for mapping files from disk to memory and its execution, relocation information, and compatibility support.  As shown in Figure \ref{fig:Windows_PE_Struct}, PE headers are followed by section headers with metadata of individual sections, including the section's size and location in disk and memory, relocation, and characteristic flags. Sections are the locations of the actual content of a PE file, which contains code, data (initialized and uninitialized), resources, and other sections with their purpose. Each section is built accordingly to meet its purpose like the \texttt{.text} section consists of codes thus made executable, while most other sections are non-executable by default. The nature of different regions of a PE malware file defines the possibilities for injecting perturbation. The headers are the most sensitive regions and, thus, have a high risk of breaking the file with any modification. Though the perturbations can be injected in sections other than the header, special care should be taken to keep the file intact after modification. The unused overlay space and the end of the file are the default locations for injecting perturbations, as they do not interfere with the file's behaviour. 

\subsection{Adversarial Evasion Attacks in Windows PE Malware}
\vspace{-2mm}
\label{subsec:adv_attcks}

\begin{figure*}
    \centering
    \includegraphics[scale=0.8]{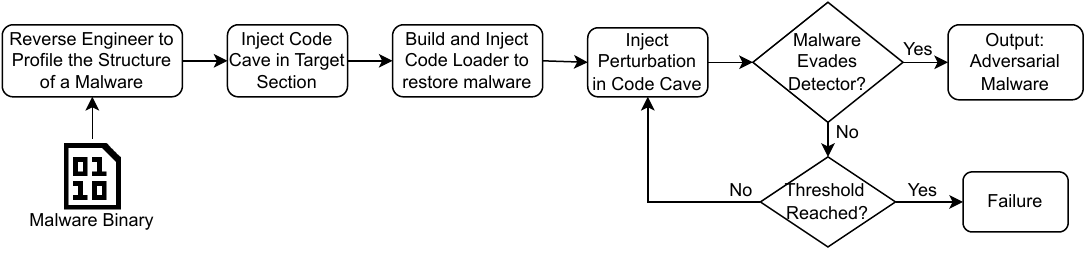}
    \vspace{-3mm}
    \caption{Flow diagram of Intra-Section code cave injection for adversarial evasion attack}
    \label{fig:flow_diagram}
   \vspace{-4mm}
\end{figure*}

Windows malware has been one of the most exploited for adversarial evasion attacks, primarily due to its wider abundance and also vulnerability. Several successful attacks use a gamut of approaches including gradient descent~\cite{kolosnjaji2018adversarial, kreuk2018adversarial}, code obfuscation~\cite{park2019generation, song2020automatic}, reinforcement learning~\cite{anderson2018learning, fang2020deepdetectnet}, Generative Adversarial Network~\cite{hu2023generating, kawai2019improved}, and others. However, for the adversarial evasion attacks on the PE file, the location for injecting perturbation is a bigger challenge than optimizing the perturbation~\cite{aryal2023exploiting}. The fragility of a binary file to some random modifications makes the location of perturbation a critical factor while carrying out such attacks on PE malware. 

Figure \ref{fig:pe_location} shows different Windows PE file locations that have been exploited to inject the adversarial perturbations. The most commonly used approach in evasion attacks for PE malware is by appending the perturbation at the end of the file~\cite{kolosnjaji2018adversarial, kreuk2018adversarial, chen2019adversarial} denoted by Region D in Figure \ref{fig:pe_location}. Injecting the perturbation at the end of the file comes with the significant advantage of not altering the functionality of a file. Since the regions at the end of the file are outside the header definition of a file, they are never executed. However, the perturbations at the end of the file are easily noticed by a malware detector. In addition, malware detectors like MalConv~\cite{raff2018malware} only consider the first 2MB of a file for detection. So, \textit{any modifications beyond the 2MB do not make any difference in malware detection}. Therefore, append attacks generally have a significantly low evasion rate~\cite{kolosnjaji2018adversarial, suciu2019exploring}.

To overcome the limitations of append attacks at the end of the file, another approach to finding the slack regions inside the file was proposed \cite{suciu2019exploring}. The slack regions are the existing empty regions(overlay) at the end of the sections created by the compiler due to a mismatch of virtual size and section alignment block size. The slack region is represented by Region C in Figure \ref{fig:pe_location}. These regions do not impact the functionality of malware but are constrained by the available size, therefore limiting the size of the perturbation that can be inserted. Work by Demetrio et al. discussed injecting perturbations in the header of malware file \cite{demetrio2019explaining} as shown by Region A in Figure \ref{fig:pe_location}. This work claimed the efficiency of perturbations in the header while elevating the risk of breaking the file, as modifications in the header are highly sensitive. 

One of the recent works injected the code caves between the sections of PE malware~\cite{yuste2022optimization}. Injection of code caves refers to introducing empty space within the file where the perturbations can be injected without impacting the file's health. The approach by Yuste et al.~\cite{yuste2022optimization}, initially adopted by ~\cite{kaspersky2005shellcoder}, takes advantage of disk space that is not mapped to memory. The code caves are injected only into the disk space but are not copied to memory. Since the approach doesn't make any difference during the execution of a program, it preserves the behavior of the adversarial file. Region B in Figure \ref{fig:pe_location} represents the inserted code cave between the \texttt{.text} and \texttt{.data} sections. Adding code caves allows the introduction of flexible perturbation amounts without any constraint. However, the unusual space and contents between the sections can easily be a red flag for malware detectors, yielding adversarial malware less effective. Table \ref{Table:adv_evasion_list} provides a comparative analysis of existing approaches in terms of different parameters, including target model impacted, perturbation space, and attack location among others. 
Based on our review and analysis of the literature, we conjecture that there are several unexplored avenues to exploit in Windows PE file structure to inject malicious perturbations yielding adversarial evasion attacks. 


\textit{In this work, we propose a novel approach to create evasion attacks that focuses on inserting code caves (and perturbations) inside the sections (intra-section) of a PE file while also maintaining the functionality and executability of the malware. As discussed later in the results section, our approach has a significantly high evasion rate (over 96\% in MalConv and 97\% in MalConv2), as compared to existing adversarial evasion approaches.}

\section{Threat Model}
\label{sec:Threat_model}
A threat model is crucial in understanding the vulnerabilities and quantifying the risk. This section outlines the key elements and considerations required to model the threat posed by adversarial malware evasion attacks. We will discuss the threats considering the attack surface and techniques, adversaries' goals, knowledge of adversaries, and their capabilities.
\begin{itemize}[leftmargin=*]
    \item \textbf{Attack Surface and Techniques: } Attack surface defines the vulnerable region that can be exploited for attacks. The test data samples are the attack surface for the AE attack. In our context, the test malware samples are modified with crafted adversarial perturbation, thus the attack surface. As for the attack technique, there can be a wide range of possibilities, from code obfuscation and structural manipulation to introducing polymorphic and metamorphic behavior. In our attacks, we adopt minor structural manipulation by injecting a code cave and a code loader, thus restoring malware to its original form.
    \item \textbf{Goals of Adversaries: }  The primary goal of an AE attack is to cause misclassification by a trained machine learning model. The goals can vary from untargeted and targeted misclassification to confidence reduction. The precise goal for our attack approach can be placed in two folds: (1) Modify malware to evade the CNN-based malware detector~\cite{raff2018malware,raff2021classifying} and (2) Maintain the functionality of malware during an attempt to evade the malware detector. 
    
    \item \textbf{Knowledge of Adversaries: } The threat posed by an adversary is proportional to the adversary's knowledge of the PE file structure, the target malware detector, and the context of the attack. Based on adversaries' knowledge, adversarial evasion attacks can be classified into three major categories: (1) Whitebox attack, where an adversary has complete knowledge about the target model; (2) Graybox attack, where an adversary has partial knowledge of the target and (3) Blackbox attack where adversary have no any knowledge of target model. Our approach uses a white-box attack where an adversary can access the target model and obtain the gradient to craft the adversarial perturbation. 
    \item \textbf{Adversarial Capabilities: } The capabilities of an attacker depend on the knowledge of adversaries; with more knowledge about the target system, the capabilities of the attacker increase and vice-versa. In this work, we can inject and modify the data in a test sample for AE attacks. The attackers can inject a perturbation inside a malware file while also being able to modify the existing contents inside a file. 
\end{itemize}
\vspace{-2mm}

\begin{figure}[!t]
    \centering
    \includegraphics[scale = 0.6]{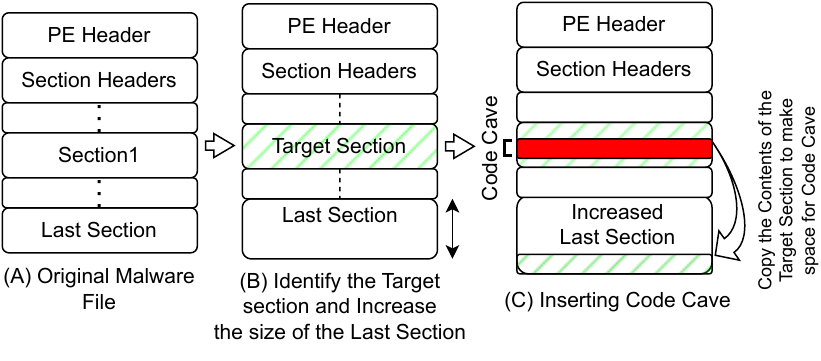}
    \vspace{-3mm}
    \caption{Inserting Intra-Section code cave in malware file.}
    \label{fig:code_caves}
    \vspace{-6mm}
\end{figure}

\section{Proposed Methodology}
\label{Sec:Methodology}
The adversarial evasion attack on Windows PE malware detectors can be divided into two tasks; (i) intra-section code cave insertion and (ii) optimizing the perturbation within those code caves. The outcome of completing these two tasks is the malware that evades a trained malware detection system while preserving the originality of the file in terms of its execution and function. Code cave insertion and perturbation optimization are insufficient to preserve the file's functionality, leading to a need for a code loader that can restore the original malware during execution. The flow diagram of our approach to generating adversarial malware samples is given in Figure \ref{fig:flow_diagram}. Detailed discussions of the steps involved in the process, starting from the raw binary to adversarial malware, are mentioned in the following subsections. 
\vspace{-2mm}
\subsection{Intra-section Code Cave Insertion}
\label{subsec: Code_Cave_Insertion}
Unlike the adversarial attacks in the image domain, the adversarial perturbations in malware must be inserted at some fixed file locations since random injection of perturbation can easily break the binary file, rendering malware useless for an adversary. The constraint of preserving a file's \textit{executability} and \textit{functionality} makes adversarial evasion attacks much more challenging in the malware binary domain. The location for injecting the perturbations depends on the binary file's structure under attack. In this work, we are exploiting the structure of Windows PE malware files to carry out an adversarial evasion attack. Since we inject an intra-section code cave in a PE file, we must consider each detail meticulously, as the modifications are highly sensitive to the file's health. In Section ~\ref{subsec:Windows_PE}, we discussed the pros and cons of injecting the perturbations in different Windows PE Malware file locations. From the detailed analysis of existing approaches (as discussed in Section \ref{sec:background}), we see significant research gaps in finding the most impactful location of inserting perturbation inside the binary PE file, which is most evasive to the detector. 

\begin{algorithm}[!t]
\caption{Algorithm for Intra-Section code cave insertion}
\begin{algorithmic}[1]
    \Statex  \textbf{Input:} Malware file $M = \{S_1, S_2, S_3,\cdots , S_n\}$, where $S_1, S_2$ are different sections and $n$ is the number of sections; target section $T \in \{S_1, S_2, S_3,.......,S_n\}$ and required code cave size $C$
    \Statex \textbf{Output:} Modified malware file $M^{'}$ with code cave and code loader in it
    \State Maximum available cave size, 
    \begin{equation*}
        P_{max} = \begin{cases}
        \text{size of } T, & \text{if T $\ne$ \texttt{.text} section}\\
        \text{size of } T \text{ - 24 bytes}, & \text{if T = \texttt{.text} section}
        \end{cases}
    \end{equation*}
    \textit{/* Last 24 bytes of .text are reserved for the code loader */}
    \State Define the starting address for a code loader, 
    $A_{Shell}$ = (starting address of \texttt{.text}) + (size of \texttt{.text}) - 24
    \If{$C > P_{max}$}
        \State Insufficient space and raise an exception
    \Else 
        \State Increase the size of the last section $S_n$ by $C$
        \State Move content of $T$ ($T_{Start}$ to $T_{End}$) of size $C$ to $S_n$
        \State Inject code loader at $A_{Shell}$ 
        \State Change program entry point (OEP) to $A_{Shell}$
    \EndIf
\end{algorithmic}
\label{Alg:Code_Cave}
\end{algorithm}

 Code cave insertion creates space inside the malware file where the adversarial perturbations can be injected. Insertion of code cave helps to combat the limited available space inside the file, which limits the amount of perturbation to be injected. Algorithm ~\ref{Alg:Code_Cave} shows our approach to inserting the code cave in original malware file $M$. We begin with reverse engineering the malware file using python libraries(PeFile and Lief) and profiling the structure of the file based on the available sections and their sizes. There could be multiple approaches to selecting the target section, but for our approach, we use the size of the section as a determining factor. If the section size is larger than 15\% of the total file size, we choose the target section in the priority order of \texttt{.text} followed by \texttt{.data} followed by the \texttt{.rdata} section. Since the \texttt{.text} section contains the code; injecting perturbation in this section is relatively easier. We start by taking an original malware file as in step (A) of Figure \ref{fig:code_caves}. We identify the target section and increase the size of the last section of the PE file to make space for contents in the code caves, as shown in step (B) of Figure \ref{fig:code_caves}. Once the target section is fixed, the contents of the target section where the code cave resides are transferred to the last section, where space has been created, as shown in step (C) of Figure \ref{fig:code_caves}. Creating the code cave in the section by moving its content breaks down the file. To preserve the originality, we inject the code loader at the end of the \texttt{.text} section, producing a modified malware $M^{'}$. The detailed explanations of the code loader are in the following subsection.

 \subsection{Preserving Functionality}
\label{subsec: Preserving_functionality}
Although the major contribution of this work involves intra-section code cave insertion and adversarial generation, numerous small details must be taken care to ensure that the file preserves its behavior. A single random operation can entirely break the process, leading to the failure of our goal of creating \textit{functional} evasive adversarial samples. The insertion of code cave alone is guaranteed to break the file as it significantly changes the sequence of code. The following subsections discuss two approaches we used to preserve the file.
\subsubsection{Code Loader Injection}
Code loaders are small chunks of code injected to perform some operations not performed by the program initially by dynamically loading the code at run time. Attackers carry out most code loader injections to force a program to perform some unintended, mostly malicious functions. In our approach, we use a code loader to restore the originality of malware during the execution of a file. Since our code loader is written on x86, it can be used to target the 32-bit architecture. As discussed in Section \ref{subsec: Code_Cave_Insertion}, the malware file, while on the disk, does have the code caves and perturbations, but its execution starts from the code loader script, which is injected in the last 24 bytes of the \texttt{.text} section. Since the program's entry point is changed to the start of the code loader, the injected code snippet restores the original structure of the malware using the Algorithm \ref{Alg:codeloader}, returning the complete original functionality to a malware file. 

\begin{algorithm}[!t]
    \caption{Operation of Code Loader on modified malware $M^{'}$}
    \begin{algorithmic}[1]
        \Statex \textbf{Input:} Starting address of code cave in target section $T$, ($T_{start}$); starting address of code cave contents located at the last section $(S_{n})$, ($S_{start}$); the size of code cave ($C$); the original entry point of the program ($OEP$) 
        \Statex \textbf{Output:} Execution of Original Malware ($M$) with control sequence starting from original entry point ($OEP$) 
        \For{$i \gets 0$ to $C$}
            \State $T\left[T_{{start}}\; \texttt{+}\; i\right]$ = $S_{n}\left[S_{start} \; \texttt{+}\; i\right]$
            \State $S_{n}\left[S_{\texttt{start}} \; \texttt{+}\; i\right]$ = `\textbackslash x00'
        \EndFor
        \State \textbf{Jump to} $OEP$
    \end{algorithmic}
    \label{Alg:codeloader}
\end{algorithm} 

Algorithm ~\ref{Alg:codeloader} shows the workflow of the code loader. The execution of the modified malware file $M^{'}$ begins with the code loader since the program's entry point is modified to its address. The code loader starts by copying back the original content of the target section to its original place from the last section, where it was copied during code cave insertion. The code cave is removed from the malware file by overwriting it with its actual content while removing the content from its temporary location in the last section. On completion of the code loader execution, the flow of the program is transferred to the original entry point of the malware $M$. The injected code loader introduces the capability to restructure the file to its original form during the execution, thus preserving the malware behavior. 

\subsubsection{Modifying Flags}
Different flags provide the attributes of the Windows PE file and its sections. We modified a few of these flags to preserve the file during our approach or to ease our process. Since the operation of our injected code loader deals with the static addresses, we disabled a file's base relocation to ensure the file is loaded at the preferred base address. To disable the base relocation, we set the IMAGE\_FILE\_RELOCS\_STRIPPED flag with $0x0001$ in the \textit{Characteristics} field of file headers. 

As the code loader modifies sections, we had to change different section's original permission. For example, our target section where code cave is placed needs permission to write as the content is written to it during execution, while the last section needs read access as it is the source of content. In addition, the section where the code loader resides should have permission to execute the code. The flags IMAGE\_SCN \_MEM\_EXECUTE, IMAGE\_SCN\_MEM\_READ, and 
IMAGE\_SCN\_MEM\_WRITE are modified for execute, read, and write permission based on the needs for each section~\cite{PEFormat48online}. 
\vspace{-2mm}
\subsection{Generating Adversarial Perturbations}
\label{subsec:Generating_adversarial}

\begin{algorithm}[!t]
    \caption{Adversarial generation using Gradient Descent}
    \label{Alg:Adv_generation}
    \begin{algorithmic}[1]
    \Statex \textbf{Input:} $x_0$, the input malware ; $q$, the maximum number of bytes that can be injected(Cave Size); $T$, the maximum number of attack iterations; $T_{Start}$, the starting address of code caves.
    \Statex \textbf{Output:} $x^{'}$ the adversarial malware sample.
    \State Set $x$ = $x_0$.
    \State Initialize perturbation bytes $q$ in $x$.
    \State Initialize the iteration counter $t$=0.
    \Repeat 
        \State Increase the iteration counter $t$ = $t$ + 1
        \For{$p$ = 1, ......,$q$}
            \State Set $j$ = $p$ + $T_{Start}$
            \State Compute the gradient $w_j$ = $-\nabla_{\phi}$${\left(x_{j}\right)}$
            \State Set $n_j$ = $w_{j}$ \textbf{\big /} $||w_{j}||_2$
            \For {i = 0, ......., 255}
                \State Compute $S_i$ = $n_{j}^{T}$ ${\left(m_{i} - z_{j}\right)}$
                \State Compute $d_i$ = $||m_{i} - \left(z_{j}\; \texttt{+} \;s_{i}.n_{j}\right)||_2$
            \EndFor
            \State {Set $x_{j}$ to $\arg \min_{i:s_{i}>0} d_{i}$.}
        \EndFor
    \Until {$f\{x\}$ $<$ 0.5 or $t$ $\geq$ $T$}
    \State return $x^{'}$
    \end{algorithmic}
\end{algorithm}

\begin{algorithm}
    \caption{Adversarial generation using FGSM}
    \label{Alg:Adv_FGSM}
    \begin{algorithmic}[1]
        \Statex \textbf{Input:} $x_0$, the input malware ; $q$, the maximum number of bytes that can be injected(Cave Size); $T$, the maximum number of attack iterations; $T_{Start}$, the starting address of code caves; $y$, the target label.
        \Statex \textbf{Output:} $x^{'}$ the adversarial malware sample.
        \State Set $x$ = $x_{0}$
        \State $x^{\text{payload}}$ $\sim$ $U\left(0, N-1\right)^{q}$
        \State ${z}^{\text{payload}}$, $z$, $\tilde{z}^{\text{payload}}$ $\leftarrow$ $M\left(x^{\text{payload}}\right)$, $M\left(x\right)$, $z^{\text{payload}}$
        \State $\tilde{z}^{\text{new}}$ $\leftarrow$ $\left[z ; \tilde{z}^{\text{payload}}\right]$
        \While{$g_{\theta}\left(\tilde{z}^{\text{new}}\right)$ $>$ 0.5}
            \State $\tilde{z}^{\text{payload}}$ $\leftarrow$ $\tilde{z}^{\text{payload}}$ - $\epsilon$ $\cdot$ $\operatorname{sign}\left(\nabla_{z}\bar{\ell}^{*} \left(\tilde{z}^{\text{new}},\tilde{y}; \theta \right) \right)$
            \State $\tilde{z}^{\text{new}}$ $\leftarrow$ $\left[z ; \tilde{z}^{\text{payload}}\right]$
        \EndWhile
        \For {$i$ $\leftarrow$ $0$ to len($x$)}
            \State $x^{'}$ $\leftarrow$ $\arg \min _j d\left(\tilde{z}_i^{\text{new}}, M_{j}\right)$
        \EndFor
        \State return $x^{'}$
    \end{algorithmic}
\end{algorithm}

At this point, we already have malware binaries with code cave and code loader. Any modifications made inside the code cave will be erased during execution with the help of a code loader. Therefore, we can inject anything in the code cave without worrying about breaking the file. Since we have created a placeholder (code cave) for our adversarial perturbation, the job is now to inject the optimal perturbations that cause to bypass the malware detectors. 

To generate adversarial binaries, we adopted and modified Algorithm \ref{Alg:Adv_generation}  proposed by Kolosnjaji et al. \cite{kolosnjaji2018adversarial} and Algorithm \ref{Alg:Adv_FGSM} proposed by Goodfellow et al.\cite{goodfellow2014explaining} and later adopted by Kreuk et al.~\cite{kreuk2018deceiving}, to inject perturbations in code caves rather than appending at the end of the file. The given algorithms carefully select perturbation bytes for code cave such that the malware binary $x_{0}$ transforms into an adversarial malware binary $x^{'}$. The size of code cave, $q$ determines the maximum number of bytes that this attack can inject. 

In Algorithm ~\ref{Alg:Adv_generation}, the gradient descent algorithm is used to optimize the perturbation bytes in this approach. The gradient of an objective function $f$ cannot be computed directly with respect to the padding byte as the MalConv architecture~\cite{raff2018malware} is not differentiable due to the presence of an embedding layer that maps each input byte $x_{j}$ to an 8-dimensional vector $z_{j}$ = $\phi\left(x_{j}\right)$. The non-differentiability of MalConv is resolved by first computing the negative gradient of $f$ with respect to embedded representation $z_{j}$, denoted as $w_{j}$ = $-\nabla_{\phi}$ $\left(x_{j}\right) \in \mathbb{R}^{8}$. The line $g_{j}(\eta)=z_{j}\; \texttt{+}\;\eta n_{j}$ is defined with normalized gradient descent $n_{j}$ = $w_{j}$ \textbf{\big /} $||w_{j}||_{2}$. The line is parallel with $w_{j}$ while passing through $z_{j}$, and $\eta$ gives its geometric locus. The perturbation byte $x_{j}$ is replaced with the corresponding embedded byte $m_{i}$ closest to the line $g_{j}$ if the projection on the line is aligned with $n_{j}$. The distance between embedded byte $m_{i}$ and the line $g_{j}$ is computed as $d_{i}$ = $||m_{i} - \left(z_{j}\; \texttt{+} \;s_{i}.n_{j}\right)||_{2}$.

Algorithm \ref{Alg:Adv_FGSM} provides a simple and fast way to generate adversarial examples. This algorithm corresponds to the Fast Gradient Sign Method (FGSM) approach of adversarial generation proposed by Goodfellow et al.~\cite{goodfellow2014explaining}. Similar to the Algorithm \ref{Alg:Adv_generation}, the input file is represented in the embedded domain $z \in Z^{*}$, which is modified to its perturbed version $\tilde{z}$ = $z$ + $\delta$. Here $f_{\theta}$ is the target malware classifier with the fixed parameters $\theta$. To generate an adversarial malware, the algorithm considers the differentiable loss function $\bar{\ell}$, max\_norm $p$ = $\infty$, and solves the problem $\tilde{z}$ = $z$ - $\epsilon \cdot \operatorname{sign} \left(\nabla_{\tilde{z}} \bar{\ell} \left(\tilde{z}, \tilde{y};\theta\right)\right)$, where $\epsilon$ is the adversarial strength. On getting the $\tilde{z}$, the new adversarial binary file $x^{'}$ is reconstructed by mapping each $z_{i}$ to the closest neighbour in the embedding matrix $M$. In place of appending perturbations at the end of the file, our approach embeds the perturbations only in the code caves, thus keeping the functionality of the file intact. 

\section{Experimental Results}
\label{Sec: Experimental_results}
\subsection{Malware Dataset}
\label{subsec:Dataset}
Our experiments involve two disjoint datasets of malicious and benign files to evaluate the effectiveness of different adversarial attacks that we orchestrated on Windows PE files. The first set comprises 6000 Windows malware binaries in 32-bit PE format obtained from a public academic repository by VirusTotal~\cite{VirusTot78:online}. This collection has \textit{eight} different malware families, including Adware (382), Downloader (897), Spyware (171), Trojan (1600), Worm (435), Virus (373), Ransomware (261), Dropper (947), and others (934) as obtained using VirusTotal public API by submitting the malware hash.
Trojan is the most common family as most of the contemporary malware have their covert behavior of disguising the user or malware detector mechanism. As it can be noted, our experiments and adversarial attacks were performed on a diverse malware family set.
The second dataset consists of 1000 Windows benign files of 32-bit PE format obtained from Kaggle~\cite{iosifach97:online}. We label all the malicious files as malware while benign files as goodwares. We do not perform feature extraction on our dataset as we use the whole executable to train the malware detector and carry out adversarial attacks on files. We verified the maliciousness of our dataset using our trained model MalConv \cite{raff2018malware} and from public malware detector VirusTotal~\cite{VirusTot78:online}. We used 80\% of the data for training the malware detector and the rest 20\% for validating our model and performing adversarial attacks. In our experimental analysis of \textit{intra-section} code cave injection attacks, out of the 1000 malware PE files, 150 malware were \textit{fit} for inserting perturbation into \texttt{.text} section, 260 for \texttt{.data} section attack, 145 for the \texttt{.rdata} section attack, and 500 malware for baseline append attacks. The malware encountered in various attacks may not always be distinct samples; there is potential for overlap in malware samples across different attacks when one or more sections surpass our specified threshold size for perturbation injection.

The number of malware considered for attacks in different sections varied based on the availability of space (15\% of the malware file size) inside the sections to fit in a code cave. As our approach deals with reverse engineering malware's structure and choosing the target section for injecting a code cave, we did not attack the packed malware in this work.

\begin{table}[]
\caption{Average section sizes in a PE malware samples}
\begin{tabular}{c|lllll|lllllllll}
\cline{2-6}
\multicolumn{1}{l|}{\textbf{}}                             & \multicolumn{5}{c|}{\textbf{Avg. Section Size Compared to PE File Size}}                                                                                         & \multicolumn{4}{c}{\textbf{}}                 & \multicolumn{2}{c}{\textbf{}} & \multirow{2}{*}{\textbf{}} & \multirow{2}{*}{\textbf{}} & \multicolumn{1}{c}{\multirow{2}{*}{\textbf{}}} \\ \cline{2-6}
                                                           & \multicolumn{1}{l|}{\textbf{}}       & \multicolumn{1}{l|}{\textbf{(0-5)\%}} & \multicolumn{1}{l|}{\textbf{(5-10)\%}} & \multicolumn{1}{l|}{\textbf{(10-15)\%}} & \textbf{15\%+} & \textbf{} & \textbf{} & \textbf{} & \textbf{} & \textbf{}     & \textbf{}     &                            &                            & \multicolumn{1}{c}{}                           \\ \cline{1-6}
\multicolumn{1}{|c|}{\multirow{5}{*}{\rotatebox[origin=c]{90}{\textbf{PE Sections}}}} & \multicolumn{1}{l|}{\textbf{.text}}  & \multicolumn{1}{l|}{32.80\%}          & \multicolumn{1}{l|}{10.14\%}           & \multicolumn{1}{l|}{19.90\%}            & 37.12\%        &           &           &           &           &               &               &                            &                            &                                                \\ \cline{2-6}
\multicolumn{1}{|c|}{}                                     & \multicolumn{1}{l|}{\textbf{.rdata}} & \multicolumn{1}{l|}{50.77\%}          & \multicolumn{1}{l|}{37.56\%}           & \multicolumn{1}{l|}{4.46\%}             & 7.17\%         &           &           &           &           &               &               &                            &                            &                                                \\ \cline{2-6}
\multicolumn{1}{|c|}{}                                     & \multicolumn{1}{l|}{\textbf{.data}}  & \multicolumn{1}{l|}{61.89\%}          & \multicolumn{1}{l|}{23.57\%}           & \multicolumn{1}{l|}{5.21\%}             & 9.29\%         &           &           &           &           &               &               &                            &                            &                                                \\ \cline{2-6}
\multicolumn{1}{|c|}{}                                     & \multicolumn{1}{l|}{\textbf{.rsrc}}  & \multicolumn{1}{l|}{46.61\%}          & \multicolumn{1}{l|}{9.61\%}            & \multicolumn{1}{l|}{6.60\%}             & 37.15\%        &           &           &           &           &               &               &                            &                            &                                                \\ \cline{2-6}
\multicolumn{1}{|c|}{}                                     & \multicolumn{1}{l|}{\textbf{.reloc}} & \multicolumn{1}{l|}{75.39\%}          & \multicolumn{1}{l|}{12.30\%}           & \multicolumn{1}{l|}{11.71\%}            & 0.49\%         &           &           &           &           &               &               &                            &                            &                                                \\ \cline{1-6}
\end{tabular}
\label{Tab:Section_Sizes}
\vspace{-3mm}
\end{table}

Another important aspect of our experiment's Windows PE Malware dataset is the presence and the sizes of different sections within a malware file. This is critical as the size of sections limits the size of code caves that can be inserted, which in turn becomes a bottleneck to the perturbation volume. This requires a comprehensive analysis of the sections in a malware sample before proceeding with the code cave insertion. We analyzed the structure of 6000 PE malware and deduced a few conclusions as follows.

\begin{itemize}[leftmargin=*, topsep=0pt]
    \item In our PE dataset, the \texttt{.text} section, the \texttt{.rdata} section, the \texttt{.data} section, the \texttt{.rsrc} section, and the \texttt{.reloc} section are present in 64\%, 49\%, 62\%, 69\%, and 22\% of the total malware samples, respectively. As we examined the sections based on their names, any malware featuring obfuscated section names was excluded from our approach.
    \item  The \texttt{.text}, \texttt{.rdata}, \texttt{.data}, and the \texttt{.rsrc} are the most likely hosts for the code caves. However, the presence of a particular section is insufficient to guarantee the injection of the code caves, as the size of the section is also critical. 
\end{itemize}

In our work, we injected the code cave in a single section, which demands our target section be large enough to have a code cave in it. We have kept the perturbation threshold to up to 15\% of the size of a malware file, meaning that the size of the target section should be at least 15\% of the size of the file. 

Table \ref{Tab:Section_Sizes} illustrates the average size of each individual section in our Windows PE Malware dataset.
Our analysis shows that 32.80\% of total malware samples have a \texttt{.text} section size less than 5\% of the total malware file size. Similarly, we can observe that only 37\% of the total malware samples have a \texttt{.text} or a \texttt{.rsrc} section occupying more than 15\% of the total malware size. Additionally, in less than 10\% of total malware samples, a \texttt{.rdata}, a \texttt{.data}, and a \texttt{.reloc} sections occupy more than 15\% of the malware file size. The availability of the section space inside Windows PE malware resulted in limiting the perturbation size to  15\% of the size of a malware file. On observing the statistics presented in Table \ref{Tab:Section_Sizes},  \texttt{.text}, \texttt{.rdata}, \texttt{.data}, and \texttt{.rsrc} are more natural targets for injecting the code cave and adversarial perturbations. The \texttt{.rsrc} section is present towards the end of the file, often outside the 2MB bound; and hence was not considered as our target for code cave in this experiment. The given statistics of section sizes in malware samples provide insight into choosing the target section to inject the code caves.

\subsection{Experimental Setup}
\label{subsec:setup}
As the experiment is built up in different phases, it requires different environments depending on the need. Since the experiments deal with binary malware, all our working environments are built in an isolated box. We used an isolated Ubuntu box to reverse engineer a malware binary and to inject the code loader and code cave. Reverse engineering deals with analyzing the structure of Windows PE malware and determining a suitable location for injecting the code cave. Python libraries lief~\cite{LIEF47:online} and PEFile~\cite{pefile·P79:online} are used to reverse engineer malware. Hex editors 010 Editor~\cite{010Edito75:online} and Ghidra~\cite{Ghidra12:online} were used to edit and verify the content of the binary file across the memory. An isolated Windows 10 virtual box was used to execute the malware file and verify its functionality after modification. The training of malware detectors and an adversarial evasion attack on them were carried out using the Pytorch library on Nvidia GPU A100 machines with 40 cores, 128GB RAM, and 500GB disk space. We used SecML Malware~\cite{demetrio2021secmlmalware} to implement the FGSM attack with some modifications.

Our experiments were initiated by training the MalConv~\cite{raff2018malware} and MalConv2~\cite{raff2021classifying} models with 6000 malware binaries obtained from VirusTotal and 1000 benign binaries obtained from Kaggle. We defined our train, test, and validation set by splitting the randomized data into a ratio of 80:10:10. Upon completion of training, our MalConv model demonstrated a detection accuracy of 94\%, whereas MalConv2 achieved an even higher accuracy of 98\% in identifying malware within our dataset. The approach to carrying out the evasion attack started manually with reverse engineering a malware file, studying its structure and execution behavior using hex editor tools. It was followed by automating the process using Python libraries.  


\begin{figure*}[!t]
    \centering
    \begin{subfigure}{0.45\textwidth}
        \includegraphics[width=\linewidth]{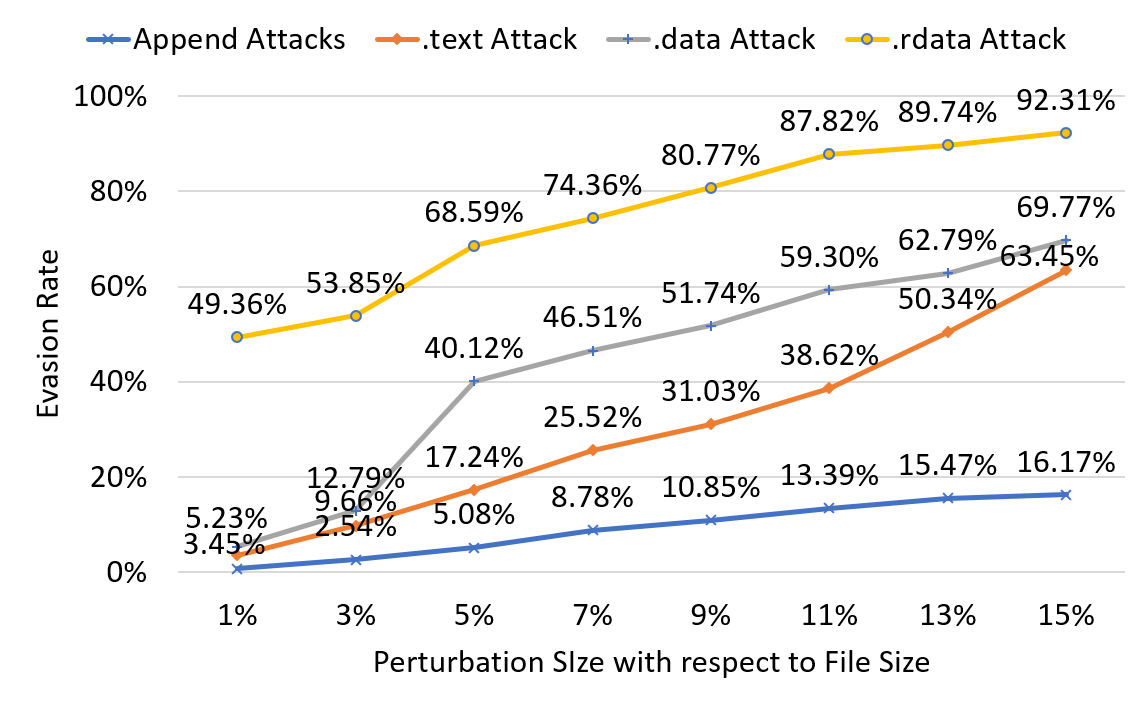} 
        \vspace{-3mm}
         \caption{Attack against MalConv}
        \label{fig:evasionratemalconv}   
    \end{subfigure}
    \hfill
    \begin{subfigure}{0.45\textwidth}
         \includegraphics[width=\linewidth]{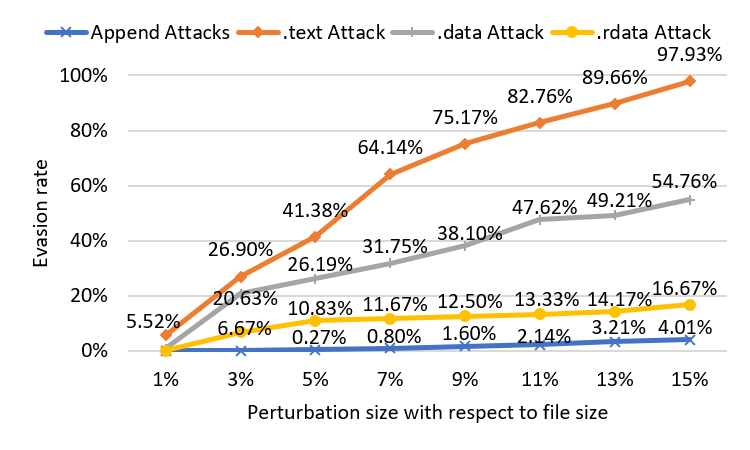}
         \vspace{-3mm}
         \caption{Attack against MalConv2}
         \label{fig:evasionratemalconv2}
    \end{subfigure}
    \vspace{-2.5mm}
    \caption{Evasion rate of \textit{intra-section} attacks in \texttt{.text}, \texttt{.data}, and \texttt{.rdata} PE file location with the baseline append attacks using gradient descent}
    \label{fig:evasionrate}
\end{figure*}

\begin{figure*}
    \centering
    \begin{subfigure}{0.45\textwidth}
        \centering \includegraphics[width=\linewidth]{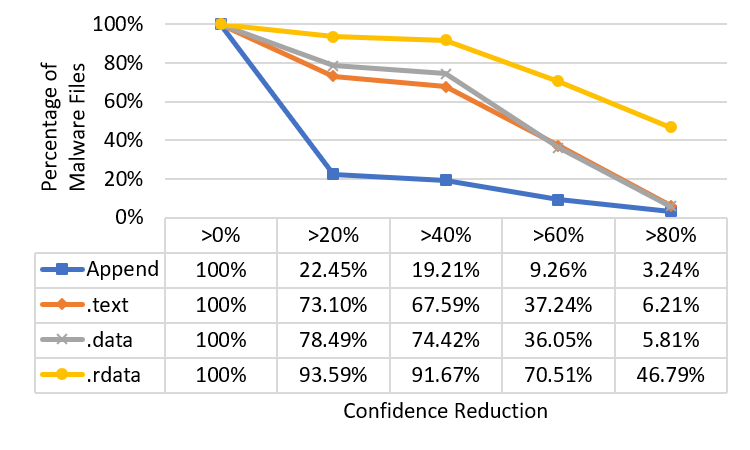}
        \caption{Confidence reduction for MalConv}
        \label{fig:confid_reduction_M1}
    \end{subfigure}
    \hfill
    \begin{subfigure}{0.45\textwidth}
        \includegraphics[width=\linewidth]{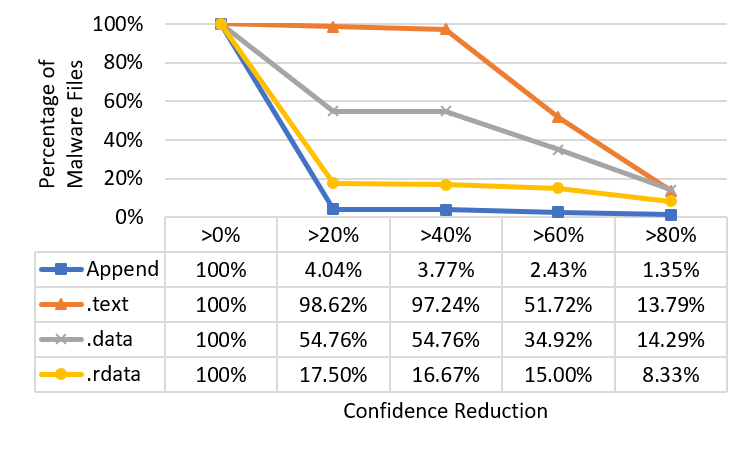}
        \caption{Confidence reduction for MalConv2}
    \label{fig:confid_reduction_M2}
    \end{subfigure}
    \vspace{-2.5mm}
    \caption{Confidence reduction after injecting 15\% perturbation in different PE file sections using gradient descent.}
    \label{fig:confid_reduction}
   \vspace{-2mm} 
\end{figure*}
\vspace{-2mm}
\subsection{Results and Analysis}
\label{subsec: results}

Here, we will address the following research questions:
\begin{itemize}[leftmargin = *, itemsep= 0pt, topsep=0pt]
    \item Can perturbations be injected within the sections rather than at the end of the file without impacting the functionality?
    \item What would be the performance of injecting perturbations within the sections compared to the end of the file?
    \item What would be the relation between perturbation size and the adversarial malware sample's evasion rate?
    \item How will adversarial attacks impact the confidence of the malware detector?
\end{itemize}

 Since the code cave is injected following the Algorithm \ref{Alg:Code_Cave}, the analysis will be carried out to measure the effectiveness of adversarial perturbation injected on those code caves. The experiment takes two major approaches to generate adversarial perturbations: Gradient descent and FGSM.

 \subsubsection{Adversarial Perturbation using Gradient Descent}
Figure \ref{fig:evasionrate} illustrates the comparative analysis of evasion rate against different attack approaches conducted in our experiments using gradient descent. Figures \ref{fig:evasionratemalconv} and \ref{fig:evasionratemalconv2} show the evasion rate against MalConv and MalConv2, respectively. 
To benchmark our proposed \textit{intra-section} approach, we first performed an append attack (similar to the ones proposed by Kolosnjaji et al.~\cite{kolosnjaji2018adversarial}) by adding the perturbation at the end of the file. For the MalConv model, these attacks achieved an evasion rate as high as 16.17\%, as compared to only 4.01\%  for MalConv2 as shown by the blue line in Figure \ref{fig:evasionrate}, with a significantly lower percentage of malware successfully evading the model. These results conflict with the original evasion obtained by Kolosnjaji et al.~\cite{kolosnjaji2018adversarial} as they claimed to achieve an evasion rate of around 60\% for MalConv. The discrepancy could be due to the higher number of iterations for their attack. Our experiments were carried out at a maximum of three iterations compared to twenty iterations in \cite{kolosnjaji2018adversarial}. Additionally, the initialization of the bytes plays a critical role in malware evasion, which is also not mentioned \cite{kolosnjaji2018adversarial}. 
Initialization of bytes with value 0 did not evade any malware sample, a phenomenon identified by another work as well~\cite{suciu2019exploring}. Note that evasion is defined as a condition if the confidence of the malware detector drops below 50\% for a particular malware file. Append attacks did not impact the malware with a size larger than 2MB, as the MalConv detector only takes the first 2MB of a file as its input. However, the limitation associated with input size is addressed in the MalConv2 architecture. Clearly, MalConv2 exhibited greater resilience to append attacks. As shown in both Figures \ref{fig:evasionratemalconv} and \ref{fig:evasionratemalconv2}, the evasion rate increased gradually with an increase in the size of the perturbation, showing a linear relationship between the evasion rate and perturbation size.

Next, we attacked the \texttt{.text} section by injecting a code cave and a code loader. This evasion attack is initiated by listing the malware from our dataset with a \texttt{.text} section large enough to hold maximum perturbation, i.e., 15\% of malware file size. We then tested the malware's functionality after injecting a code cave and a code loader. The execution on the Windows box and debugger verified the originality of the malware's behavior. The adversarial perturbations, to be injected in a code cave, were generated using a gradient-based approach as discussed in Algorithm ~\ref{Alg:Adv_generation}. Attacking the \texttt{.text} section gave an evasion rate as high as 63.45\% and 97.93\% against MalConv and MalConv2, respectively, with 15\% perturbation. Figure \ref{fig:evasionratemalconv} shows that the attack on the \texttt{.text} section (indicated by the red line) surpasses the performance of the append attack against MalConv. Furthermore, the similar attack exhibited a significantly high evasion rate over all other attacks when targeting MalConv2, as shown in Figure \ref{fig:evasionratemalconv2}.

To further verify our \textit{intra-section} approach, we performed attacks on two additional PE file sections, \texttt{.data} and \texttt{.rdata}. The attacks on the \texttt{.data} section yielded an evasion rate of up to 69.77\% against MalConv and 54.76\% against MalConv2 with 15\% perturbation. The attacks targeted at the \texttt{.data} section demonstrated higher evasion rate compared to the \texttt{.text} and the append attacks when tested against MalConv. Further, the  \texttt{.data} attacks surpassed the attacks on the \texttt{.rdata} sections and append attacks when evaluated against MalConv2. 
The attack on the \texttt{.rdata} section achieved a maximum evasion rate of 92.31\% against the MalConv, significantly outperforming the attacks in all other sections, meaning 92.31\% of the adversarially created malware was undetected by the MalConv, as shown by yellow line in Figure \ref{fig:evasionratemalconv}.
Interestingly, the attacks on the \texttt{.rdata} section produced an evasion rate of 49.36\% with the addition of just 1\% adversarial perturbation against MalConv. On the contrary, the attack on the \texttt{.rdata} section could only achieve an evasion rate of 16.67\% against MalConv2, just outperforming the append attacks as shown by the Figure \ref{fig:evasionratemalconv2}.


The results in Figure \ref{fig:evasionrate} show interesting differences while conducting attacks on the same environment but targeting different models. While using MalConv as a target, injecting the perturbation in the \texttt{.rdata} section results in the highest evasion rate, while perturbations in the \texttt{.text} section leads to the lowest evasion rate among the intra-section attacks. However, on targeting MalConv2, the \texttt{.text} section perturbations cause the highest evasion and the \texttt{.rdata} section causing the lowest. These variations are due to how these models prioritize different sections in their decision-making process.




Although \textit{evasion rate} is an important parameter to understand the effectiveness of the adversarial attack, to reinforce our \textit{intra-section} approach and experimental results, we will also discuss \textit{confidence reduction} metric. 
It's worth noting that there could be instances of malware for which MalConv's confidence drops from 100\% to 51\% following an adversarial attack but is still not considered evasion. 
Figure \ref{fig:confid_reduction} shows the impact of adversarial evasion attacks on the confidence reduction of MalConv and MalConv2 models in detecting the malware. Even in the \textit{confidence reduction} metric, we can see the append attack performs the worst as compared to intra-section injection. For MalConv, attacks on \texttt{.text} and \texttt{.data} sections show a comparable confidence reduction of more than 80\% for around 6\% of adversarial malware files as shown in Figure \ref{fig:confid_reduction_M1}. Further, the attack on \texttt{.rdata} section outperforms other attacks as more than 46\% of adversarial malware PE files get a confidence reduction of 80\% or more against MalConv. Similarly, with MalConv2 as target model (Figure \ref{fig:confid_reduction_M2}), the attacks on the \texttt{.text} and \texttt{.data} sections result in a confidence reduction of over 80\% for approximately 14\% of malware, which is higher compared to MalConv as the target. These results clearly reflect our approach outperforms the append attacks while also highlighting an interesting difference in the performance of the attacks while targeting different PE file sections using a gradient descent approach. The results indicate that MalConv2 exhibits greater resilience against append attacks and attacks on the \texttt{.rdata} section while showing higher vulnerability when targeted in the \texttt{.text} section. These results successfully demonstrate the significance of the locations for injecting adversarial perturbations within a malware file.

\subsubsection{Adversarial Perturbation using FGSM}
\label{Subsec:Adv_Perturbation_FGSM}

\begin{table}[t]
\centering
\caption{Evasion rate of intra-section attacks in \texttt{.text}, \texttt{.data}, and \texttt{.rdata} location with the baseline append attacks against MalConv using FGSM}
\begin{tabular}{|lllll|llllllllll}
\cline{1-5}
\multicolumn{1}{|c||}{\multirow{2}{*}{\textbf{Attacks}}} & \multicolumn{4}{c|}{\textbf{Perturbation   Volume in Bytes}}                                                                  & \multicolumn{1}{c}{} & \multicolumn{1}{c}{\textbf{}} &           &           &           & \multicolumn{1}{c}{\textbf{}} &           & \textbf{} & \textbf{} & \multicolumn{1}{c}{\textbf{}} \\ \cline{2-5}
\multicolumn{1}{|c||}{}                                  & \multicolumn{1}{l|}{\textbf{2048}} & \multicolumn{1}{l|}{\textbf{4096}} & \multicolumn{1}{l|}{\textbf{8192}} & \textbf{16384} & \textbf{}            & \textbf{}                     & \textbf{} & \textbf{} & \textbf{} & \textbf{}                     & \textbf{} & \textbf{} &           & \textbf{}                     \\ \cline{1-5}
\noalign{\vskip\doublerulesep
         \vskip-\arrayrulewidth}
\cline{1-5}
\multicolumn{5}{|c|}{\textbf{Epsilon 1.0}}                                                                                                                                              &                      &                               &           &           &           &                               &           &           &           &                               \\ \cline{1-5}
\multicolumn{1}{|l|}{\textbf{Append}}                   & \multicolumn{1}{l|}{35.23\%}       & \multicolumn{1}{l|}{35.23\%}       & \multicolumn{1}{l|}{35.23\%}       & 35.23\%        &                      &                               &           &           &           &                               &           &           &           &                               \\ \cline{1-5}
\multicolumn{1}{|l|}{\textbf{.text}}                    & \multicolumn{1}{l|}{12.60\%}       & \multicolumn{1}{l|}{12.60\%}       & \multicolumn{1}{l|}{12.60\%}       & 14.17\%        &                      &                               &           &           &           &                               &           &           &           &                               \\ \cline{1-5}
\multicolumn{1}{|l|}{\textbf{.data}}                    & \multicolumn{1}{l|}{25.00\%}       & \multicolumn{1}{l|}{48.24\%}       & \multicolumn{1}{l|}{50.59\%}       & 45.18\%        &                      &                               &           &           &           &                               &           &           &           &                               \\ \cline{1-5}
\multicolumn{1}{|l|}{\textbf{.rdata}}                   & \multicolumn{1}{l|}{66.36\%}       & \multicolumn{1}{l|}{86.92\%}       & \multicolumn{1}{l|}{93.46\%}       & 96.26\%        &                      &                               &           &           &           &                               &           &           &           &                               \\ \cline{1-5}
\noalign{\vskip\doublerulesep
         \vskip-\arrayrulewidth}
\cline{1-5}
\multicolumn{5}{|c|}{\textbf{Epsilon 0.75}}                                                                                                                                             &                      &                               &           &           &           &                               &           &           &           &                               \\ \cline{1-5}
\multicolumn{1}{|l|}{\textbf{Append}}                   & \multicolumn{1}{l|}{39.57\%}       & \multicolumn{1}{l|}{39.57\%}       & \multicolumn{1}{l|}{39.57\%}       & 39.57\%        &                      &                               &           &           &           &                               &           &           &           &                               \\ \cline{1-5}
\multicolumn{1}{|l|}{\textbf{.text}}                    & \multicolumn{1}{l|}{11.81\%}       & \multicolumn{1}{l|}{11.81\%}       & \multicolumn{1}{l|}{11.81\%}       & 13.34\%        &                      &                               &           &           &           &                               &           &           &           &                               \\ \cline{1-5}
\multicolumn{1}{|l|}{\textbf{.data}}                    & \multicolumn{1}{l|}{26.32\%}       & \multicolumn{1}{l|}{48.68\%}       & \multicolumn{1}{l|}{49.56\%}       & 61.40\%        &                      &                               &           &           &           &                               &           &           &           &                               \\ \cline{1-5}
\multicolumn{1}{|l|}{\textbf{.rdata}}                   & \multicolumn{1}{l|}{66.36\%}       & \multicolumn{1}{l|}{87.85\%}       & \multicolumn{1}{l|}{93.46\%}       & 96.26\%        &                      &                               &           &           &           &                               &           &           &           &                               \\ \cline{1-5}
\noalign{\vskip\doublerulesep
         \vskip-\arrayrulewidth}
\cline{1-5}
\multicolumn{5}{|c|}{\textbf{Epsilon 0.50}}                                                                                                                                             &                      &                               &           &           &           &                               &           &           &           &                               \\ \cline{1-5}
\multicolumn{1}{|l|}{\textbf{Append}}                   & \multicolumn{1}{l|}{40.92\%}       & \multicolumn{1}{l|}{40.92\%}       & \multicolumn{1}{l|}{40.92\%}       & 38.07\%        &                      &                               &           &           &           &                               &           &           &           &                               \\ \cline{1-5}
\multicolumn{1}{|l|}{\textbf{.text}}                    & \multicolumn{1}{l|}{17.32\%}       & \multicolumn{1}{l|}{17.32\%}       & \multicolumn{1}{l|}{17.32\%}       & 18.90\%        &                      &                               &           &           &           &                               &           &           &           &                               \\ \cline{1-5}
\multicolumn{1}{|l|}{\textbf{.data}}                    & \multicolumn{1}{l|}{29.82\%}       & \multicolumn{1}{l|}{34.21\%}       & \multicolumn{1}{l|}{50.88\%}       & 63.60\%        &                      &                               &           &           &           &                               &           &           &           &                               \\ \cline{1-5}
\multicolumn{1}{|l|}{\textbf{.rdata}}                   & \multicolumn{1}{l|}{86.92\%}       & \multicolumn{1}{l|}{90.65\%}       & \multicolumn{1}{l|}{94.39\%}       & 96.26\%        &                      &                               &           &           &           &                               &           &           &           &                               \\ \cline{1-5}
\noalign{\vskip\doublerulesep
         \vskip-\arrayrulewidth}
\cline{1-5}
\multicolumn{5}{|c|}{\textbf{Epsilon 0.25}}                                                                                                                                             &                      &                               &           &           &           &                               &           &           &           &                               \\ \cline{1-5}
\multicolumn{1}{|l|}{\textbf{Append}}                   & \multicolumn{1}{l|}{19.94\%}       & \multicolumn{1}{l|}{19.94\%}       & \multicolumn{1}{l|}{23.85\%}       & 21.31\%        &                      &                               &           &           &           &                               &           &           &           &                               \\ \cline{1-5}
\multicolumn{1}{|l|}{\textbf{.text}}                    & \multicolumn{1}{l|}{14.96\%}       & \multicolumn{1}{l|}{14.96\%}       & \multicolumn{1}{l|}{14.96\%}       & 16.54\%        &                      &                               &           &           &           &                               &           &           &           &                               \\ \cline{1-5}
\multicolumn{1}{|l|}{\textbf{.data}}                    & \multicolumn{1}{l|}{28.51\%}       & \multicolumn{1}{l|}{30.26\%}       & \multicolumn{1}{l|}{38.60\%}       & 46.05\%        &                      &                               &           &           &           &                               &           &           &           &                               \\ \cline{1-5}
\multicolumn{1}{|l|}{\textbf{.rdata}}                   & \multicolumn{1}{l|}{83.18\%}       & \multicolumn{1}{l|}{89.72\%}       & \multicolumn{1}{l|}{92.52\%}       & 95.33\%        &                      &                               &           &           &           &                               &           &           &           &                               \\ \cline{1-5}
\end{tabular}
\label{Tab:FGSM_results}
\end{table}
To further test the effectiveness of our approach, we used the Fast Gradient Sign Method to generate the adversarial perturbations. We adopted the FGSM implementation from the SecML-Malware library~\cite{demetrio2021secmlmalware} to generate and inject perturbations to code cave in different locations of a malware PE file against the trained MalConv~\cite{raff2018malware} model with different values of Epsilon, $\epsilon \in [0.25, 0.50, 0.75, 1.0]$ and perturbation volumes of [2048, 4096, 8192 \& 16384] bytes. However, FGSM for MalConv2 was implemented using PyTorch due to differences in the structure of the MalConv and MalConv2 models. FGSM was implemented with norm value, $p_{norm}$ = $\infty$ and iterations up to 20. Table \ref{Tab:FGSM_results} and Table \ref{Tab:FGSM_results_malconv2} show the evasion rate of intra-section attacks in \texttt{.text}, \texttt{.data} and \texttt{.rdata}, compared to the append attacks with MalConv and MalConv2 as a target model, respectively. FGSM increased the evasion rate of append attacks up to 40.92\% compared to 16.17\% achieved by the gradient descent method against MalConv and up to 54.75\% from 4.01\% against MalConv2. The nature of FGSM helps it efficiently generate adversarial perturbation w.r.t time taken to generate and the size of perturbation. 
Furthermore, it is evident that the evasion rate of append attacks has escalated against MalConv2 in contrast to MalConv. This surge in evasion against MalConv2 can be attributed to its capacity to process larger input sizes compared to MalConv, enabling it to consider the appended perturbation beyond 2MB.

FGSM achieved the maximum evasion rate of 18.90\%, 63.60\% and 96.26\% for attacking the \texttt{.text}, \texttt{.data} and \texttt{.rdata} sections, respectively, against MalConv.
Likewise, when targeting the \texttt{.text}, \texttt{.data}, and \texttt{.rdata} sections with FGSM attacks, the maximum evasion rates against MalConv2 reached 73.86\%, 77.69\%, and 94.34\%, respectively. 
Although there are variations in the evasion rates observed between gradient descent and FGSM techniques, our approach consistently outperformed the baseline append attacks in the majority of cases. These maximum evasion rates are achieved with the value of epsilon ($\epsilon$)=0.5 against MalConv and epsilon ($\epsilon$)=1 against MalConv2. Moreover, FGSM, being a more efficient algorithm compared to gradient descent, achieved a higher evasion rate with significantly less perturbation volume.

\begin{table}[t]
\centering
\caption{Evasion rate of intra-section attacks in \texttt{.text}, \texttt{.data}, and \texttt{.rdata} location with the baseline append attacks against MalConv2 using FGSM}
\begin{tabular}{|lllll|llllllllll}
\cline{1-5}
\multicolumn{1}{|c||}{\multirow{2}{*}{\textbf{Attacks}}} & \multicolumn{4}{c|}{\textbf{Perturbation   Volume in Bytes}}                                                                  & \multicolumn{1}{c}{} & \multicolumn{1}{c}{\textbf{}} &           &           &           & \multicolumn{1}{c}{\textbf{}} &           & \textbf{} & \textbf{} & \multicolumn{1}{c}{\textbf{}} \\ \cline{2-5}
\multicolumn{1}{|c||}{}                                  & \multicolumn{1}{l|}{\textbf{2048}} & \multicolumn{1}{l|}{\textbf{4096}} & \multicolumn{1}{l|}{\textbf{8192}} & \textbf{16384} & \textbf{}            & \textbf{}                     & \textbf{} & \textbf{} & \textbf{} & \textbf{}                     & \textbf{} & \textbf{} &           & \textbf{}                     \\ \cline{1-5}
\noalign{\vskip\doublerulesep
         \vskip-\arrayrulewidth}
\cline{1-5}
\multicolumn{5}{|c|}{\textbf{Epsilon 1.0}}                                                                                                                                              &                      &                               &           &           &           &                               &           &           &           &                               \\ \cline{1-5}
\multicolumn{1}{|l|}{\textbf{Append}}                   & \multicolumn{1}{l|}{31.48\%}       & \multicolumn{1}{l|}{40.05\%}       & \multicolumn{1}{l|}{46.64\%}       & 54.75\%        &                      &                               &           &           &           &                               &           &           &           &                               \\ \cline{1-5}
\multicolumn{1}{|l|}{\textbf{.text}}                    & \multicolumn{1}{l|}{73.20\%}       & \multicolumn{1}{l|}{73.20\%}       & \multicolumn{1}{l|}{73.20\%}       & 73.86\%        &                      &                               &           &           &           &                               &           &           &           &                               \\ \cline{1-5}
\multicolumn{1}{|l|}{\textbf{.data}}                    & \multicolumn{1}{l|}{33.06\%}       & \multicolumn{1}{l|}{56.61\%}       & \multicolumn{1}{l|}{66.12\%}       & 77.69\%        &                      &                               &           &           &           &                               &           &           &           &                               \\ \cline{1-5}
\multicolumn{1}{|l|}{\textbf{.rdata}}                   & \multicolumn{1}{l|}{39.62\%}       & \multicolumn{1}{l|}{44.03\%}       & \multicolumn{1}{l|}{89.94\%}       & 94.34\%        &                      &                               &           &           &           &                               &           &           &           &                               \\ \cline{1-5}
\noalign{\vskip\doublerulesep
         \vskip-\arrayrulewidth}
\cline{1-5}
\multicolumn{5}{|c|}{\textbf{Epsilon 0.75}}                                                                                                                                             &                      &                               &           &           &           &                               &           &           &           &                               \\ \cline{1-5}
\multicolumn{1}{|l|}{\textbf{Append}}                   & \multicolumn{1}{l|}{29.40\%}       & \multicolumn{1}{l|}{37.50\%}       & \multicolumn{1}{l|}{45.37\%}       & 51.74\%        &                      &                               &           &           &           &                               &           &           &           &                               \\ \cline{1-5}
\multicolumn{1}{|l|}{\textbf{.text}}                    & \multicolumn{1}{l|}{62.09\%}       & \multicolumn{1}{l|}{62.09\%}       & \multicolumn{1}{l|}{62.09\%}       & 62.09\%        &                      &                               &           &           &           &                               &           &           &           &                               \\ \cline{1-5}
\multicolumn{1}{|l|}{\textbf{.data}}                    & \multicolumn{1}{l|}{32.64\%}       & \multicolumn{1}{l|}{47.93\%}       & \multicolumn{1}{l|}{65.29\%}       &76.86\%        &                      &                               &           &           &           &                               &           &           &           &                               \\ \cline{1-5}
\multicolumn{1}{|l|}{\textbf{.rdata}}                   & \multicolumn{1}{l|}{38.99\%}       & \multicolumn{1}{l|}{43.40\%}       & \multicolumn{1}{l|}{89.31\%}       & 93.71\%        &                      &                               &           &           &           &                               &           &           &           &                               \\ \cline{1-5}
\noalign{\vskip\doublerulesep
         \vskip-\arrayrulewidth}
\cline{1-5}
\multicolumn{5}{|c|}{\textbf{Epsilon 0.50}}                                                                                                                                             &                      &                               &           &           &           &                               &           &           &           &                               \\ \cline{1-5}
\multicolumn{1}{|l|}{\textbf{Append}}                   & \multicolumn{1}{l|}{26.16\%}       & \multicolumn{1}{l|}{35.06\%}       & \multicolumn{1}{l|}{41.90\%}       & 48.26\%        &                      &                               &           &           &           &                               &           &           &           &                               \\ \cline{1-5}
\multicolumn{1}{|l|}{\textbf{.text}}                    & \multicolumn{1}{l|}{17.65\%}       & \multicolumn{1}{l|}{17.65\%}       & \multicolumn{1}{l|}{17.65\%}       & 18.95\%        &                      &                               &           &           &           &                               &           &           &           &                               \\ \cline{1-5}
\multicolumn{1}{|l|}{\textbf{.data}}                    & \multicolumn{1}{l|}{29.75\%}       & \multicolumn{1}{l|}{40.50\%}       & \multicolumn{1}{l|}{55.79\%}       & 68.60\%        &                      &                               &           &           &           &                               &           &           &           &                               \\ \cline{1-5}
\multicolumn{1}{|l|}{\textbf{.rdata}}                   & \multicolumn{1}{l|}{38.37\%}       & \multicolumn{1}{l|}{42.14\%}       & \multicolumn{1}{l|}{88.05\%}       & 93.08\%        &                      &                               &           &           &           &                               &           &           &           &                               \\ \cline{1-5}
\noalign{\vskip\doublerulesep
         \vskip-\arrayrulewidth}
\cline{1-5}
\multicolumn{5}{|c|}{\textbf{Epsilon 0.25}}                                                                                                                                             &                      &                               &           &           &           &                               &           &           &           &                               \\ \cline{1-5}
\multicolumn{1}{|l|}{\textbf{Append}}                   & \multicolumn{1}{l|}{23.84\%}       & \multicolumn{1}{l|}{31.94\%}       & \multicolumn{1}{l|}{38.66\%}       & 45.02\%        &                      &                               &           &           &           &                               &           &           &           &                               \\ \cline{1-5}
\multicolumn{1}{|l|}{\textbf{.text}}                    & \multicolumn{1}{l|}{4.57\%}       & \multicolumn{1}{l|}{4.57\%}       & \multicolumn{1}{l|}{5.23\%}       & 7.84\%        &                      &                               &           &           &           &                               &           &           &           &                               \\ \cline{1-5}
\multicolumn{1}{|l|}{\textbf{.data}}                    & \multicolumn{1}{l|}{26.86\%}       & \multicolumn{1}{l|}{34.30\%}       & \multicolumn{1}{l|}{50.00\%}       & 65.29\%        &                      &                               &           &           &           &                               &           &           &           &                               \\ \cline{1-5}
\multicolumn{1}{|l|}{\textbf{.rdata}}                   & \multicolumn{1}{l|}{37.11\%}       & \multicolumn{1}{l|}{40.88\%}       & \multicolumn{1}{l|}{86.79\%}       & 93.08\%        &                      &                               &           &           &           &                               &           &           &           &                               \\ \cline{1-5}
\end{tabular}
\label{Tab:FGSM_results_malconv2}
\end{table}
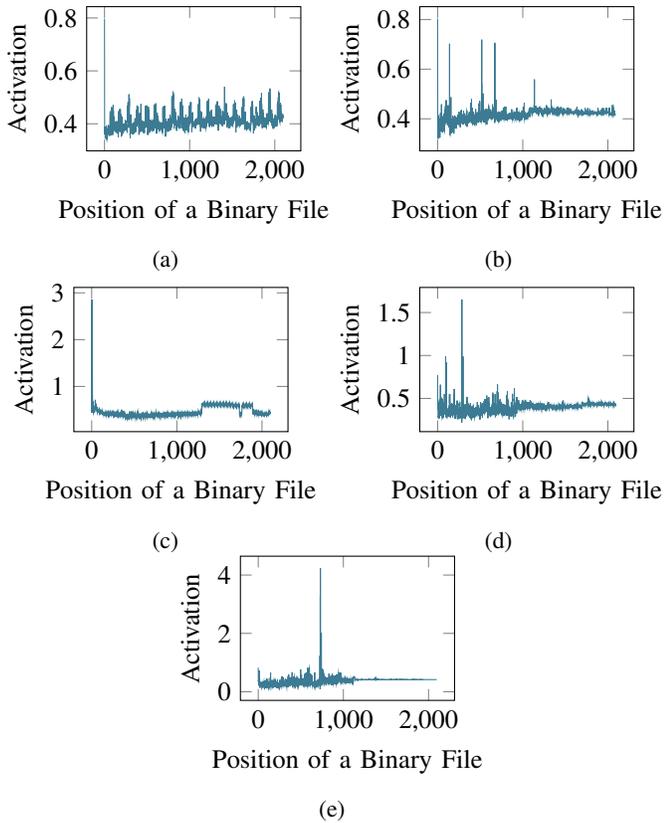
\begin{figure}[t]
    \begin{subfigure}{0.5\linewidth}
        \centering
        \pgfplotsset{compat=1.18}
        \begin{tikzpicture}
         \begin{axis}[
            width=\linewidth,
            height=3.5cm,
            minor tick style = {draw=none},
            xlabel = Position of a Binary File,
            ylabel = Activation]
            \addplot[color=black!50!cyan, smooth] table [x=x_values, y=y_values, col sep=comma]{Activation_CSV/Original_Activation_Multiplied.csv};
        \end{axis}
        \end{tikzpicture}
        \caption{}
        \label{Subfig:Activ_Original}
    \end{subfigure}%
    \begin{subfigure}{0.5\linewidth}
        \centering
        \pgfplotsset{compat=1.18}
        \begin{tikzpicture}
         \begin{axis}[
            width=\linewidth,
            height=3.5cm,
            minor tick style = {draw=none},
            xlabel = Position of a Binary File,
            ylabel = Activation]
            \addplot[color=black!50!cyan, smooth] table [x=x_values, y=y_values, col sep=comma]{Activation_CSV/Append_Activation_Multiplied.csv};
        \end{axis}
        \end{tikzpicture}
        \caption{}
        \label{Subfig:Activ_Append}
    \end{subfigure}\\
    \begin{subfigure}{0.5\linewidth}
        \centering
        \pgfplotsset{compat=1.18}
        \begin{tikzpicture}
         \begin{axis}[
            width=\linewidth,
            height=3.5cm,
            minor tick style = {draw=none},
            xlabel = Position of a Binary File,
            ylabel = Activation]
            \addplot[color=black!50!cyan, smooth] table [x=x_values, y=y_values, col sep=comma]{Activation_CSV/text_Activation_Multiplied.csv};
        \end{axis}
        \end{tikzpicture}
        \caption{}
        \label{Subfig:Activ_text}
    \end{subfigure}%
    \begin{subfigure}{0.5\linewidth}
        \centering
        \pgfplotsset{compat=1.18}
        \begin{tikzpicture}
         \begin{axis}[
            width=\linewidth,
            height=3.5cm,
            minor tick style = {draw=none},
            xlabel = Position of a Binary File,
            ylabel = Activation]
            \addplot[color=black!50!cyan, smooth] table [x=x_values, y=y_values, col sep=comma]{Activation_CSV/Data_Activation_Multiplied.csv};
        \end{axis}
        \end{tikzpicture}
        \caption{}
        \label{Subfig:Activ_data}
    \end{subfigure}\\
    \begin{subfigure}{\linewidth}
        \centering
        \pgfplotsset{compat=1.18}
        \begin{tikzpicture}
         \begin{axis}[
            width=0.5\linewidth,
            height=3.5cm,
            minor tick style = {draw=none},
            xlabel = Position of a Binary File,
            ylabel = Activation]
            \addplot[color=black!50!cyan, smooth] table [x=x_values, y=y_values, col sep=comma]{Activation_CSV/Rdata_Activation_Multiplied.csv};
        \end{axis}
        \end{tikzpicture}
        \caption{}
        \label{Subfig:Activ_rdata}
    \end{subfigure}
    \caption{Multiplied average convolutional activation of ReLU and Sigmoid for MalConv on different malware samples (a) original;  adversarial generated with (b) append attacks, (c) \texttt{.text}, (d) \texttt{.data}, (e) \texttt{.rdata} sections, respectively.}
    \label{Fig:Convolutional_Activ}
    \vspace{-6mm}
\end{figure}

Despite FGSM being more efficient, there were some uncommon behaviors observed during the \textit{intra-section} attacks. Firstly, there is no clear reason why the attack in \texttt{.text} section is underperforming as compared to other \textit{intra-section} attacks. Secondly, as observed in Tables \ref{Tab:FGSM_results} and \ref{Tab:FGSM_results_malconv2}, an increase in the perturbation size from 2048 bytes to 16384 bytes for \texttt{.text} and append attacks has no substantial difference in the evasion rate for few instances. Finally, most of the evasion was created in the first few iterations, and any number of iterations after that did not make any difference in the attack's success.

To answer some of the above behavior, we plotted a convolutional activation plot for the MalConv model as shown in Figure \ref{Fig:Convolutional_Activ}. As MalConv has two different 1-d CNN networks, it gives two different activations, ReLU and Sigmoid, which are multiplied before feeding to the max pooling layer. We plot the averaged multiplied activation of the convolutional layer while passing the original and the adversarial malware. The evasion from detection can be explained from the perspective of a shift in the model's attention, as previously mentioned by Kreuk et al.~\cite{kreuk2018deceiving}. For original malware samples, Figure \ref{Subfig:Activ_Original} shows that activation is distributed uniformly across the binary file except at the start of the file, indicating the header regions, suggesting that the header of a PE makes a significant contribution towards detection, while the rest of the regions have a similar contribution. On appending the perturbation at the end of the file, Figure \ref{Subfig:Activ_Append} shows the activation spikes in regions other than the header across the binary file. One would naturally expect those activation spikes at the end for append attacks; however, due to the variable size of the binary file in the malware sample, it is distributed across the file. We can see a difference in convolutional activation for adversarial malware samples generated by intra-section attack in \texttt{.text} section in Figure \ref{Subfig:Activ_text}. The addition of perturbation in \texttt{.text} section further enhanced the activation from the header of malware in place of creating the distraction, explaining why the perturbations in the \texttt{.text} section is not making a big impact on evading the malware. Further, in Figures \ref{Subfig:Activ_data} and \ref{Subfig:Activ_rdata}, we can observe the adversarial perturbations injected in the \texttt{.data} and \texttt{.rdata} are creating the activation much larger in comparison to the activation from any other section of malware, thus leading to high evasion rate. 

\section{Discussions and Challenges}
\label{sec:Discussion}
Our experiments have shown significant results with the proposed approach to generating adversarial malware binary using intra-section perturbations injection. Regardless of our efforts to be more general, accurate, and efficient, we can not rule out the presence of unseen biases in our test data, algorithm, or even approach. There are still open challenges and gaps that should be addressed to make our approach more compatible with real-world scenarios. In this section, we will discuss challenges faced during our research, solutions used to address them, and some limitations. 
\begin{itemize}[leftmargin = *, itemsep=0 pt]
    \item \textbf{Practicality Challenge:} Before we proceed into further discussions regarding our approach, we want to acknowledge some practical challenges in pursuing this approach at this stage. Though our approach is tested against white-box attacks, we have yet to test it in a black-box environment. However, given the transferability of perturbations, we anticipate the approach to exhibit similar efficacy in black-box scenarios with attacks on surrogate models. Additionally, it's important to acknowledge some modifications of flags and disabling the ASLR flag in the executable as another unavoidable modification that had to be carried out at the current stage of the work. 
    \item \textbf{Malware Detector:} Our approach used MalConv~\cite{raff2018malware} and MalConv2~\cite{raff2021classifying} architectures as a malware detector trained on 6000 malware and 1000 benign files. While we did attain a commendable detection accuracy of 94\% with MalConv and an even more impressive 98\% with MalConv2, it's important to acknowledge that the real-world landscape of malware detection may introduce additional challenges and factors that we cannot disregard.
    Some works have already explored and exploited the vulnerabilities of MalConv~\cite{suciu2019exploring,demetrio2019explaining}. Our focus on targeting MalConv was motivated by its widespread recognition as a standard benchmark for conducting adversarial evasion attacks on Windows PE Malware~\cite{kreuk2018adversarial, kolosnjaji2018adversarial, suciu2019exploring, demetrio2019explaining}. To our knowledge, it is the first adversarial evasion attack against the MalConv2 model. However, the value of our approach extends beyond a single model. It holds the potential for broader applicability, providing a basis for evaluating and adapting similar strategies to other commercial malware detectors. In addition, we have used a white-box approach with complete information about the malware detector. The gradient value given by the detector is used to guide the perturbation generation for adversarial malware. In practice, attackers typically operate under more constrained conditions, lacking intimate knowledge of the target malware detector.


    \item \textbf{Adversarial Generation Algorithm:} To generate the adversarial perturbation for the malware, we employed the gradient descent algorithm originally proposed by Kolosnjaji et al.~\cite{kolosnjaji2018adversarial} and Fast Gradient Sign Method by Kreuk et al.~\cite{kreuk2018deceiving}. While these algorithms yielded promising results, we encountered some intriguing phenomena that need discussion. We initialized the perturbation byte with the value '0', as proposed by Kolosnjaji et al.~\cite{kolosnjaji2018adversarial}. However, none of the attacks successfully generated adversarial malware, as mentioned by Suciu et al.~\cite{suciu2019exploring}. However, we successfully generated adversarial malware samples after initializing the code cave with the original content of the target section. We could not find any conclusive reasoning behind this behavior of the algorithm. The gradient-based approach needs the perturbation byte to be updated in each iteration, resulting in the oscillatory behavior for the convergence. This makes the gradient-based approach inefficient in terms of the size of the perturbation and time for optimization. Moreover, there are some unexplained behaviours in FGSM as well as discussed in Section \ref{Subsec:Adv_Perturbation_FGSM}. However, in this work, we are not concerned about the efficiency of our algorithm. We aim to propose a novel intra-section code cave injection approach while ensuring the malware's functionality.

    \item \textbf{Code Caves and Perturbation Size:} Code caves injection have recently been researched in adversarial evasion attacks \cite{yuste2022optimization}. 
    Clearly, code caves injection within the PE file sections has a significant evasion rate compared to append attacks. 
    Hiding the code cave inside a section makes it less evident for malware detectors as the perturbations are co-located with the regular malware contents. However, not all PE sections can hold the desired size of perturbations inside them. 
    Further, all the sections may not be large enough to hold the perturbation of the required size, and not all the malware have the section size sufficient to hold the adversarial perturbation. Evidently, our approach to intra-section attacks is limited to only those malware samples with at least one of the sections large enough to hold the 15\% perturbation relative to the file size. 

    \item \textbf{Creation and Injection of Code Loader:} The injection of the code loader is a critical element of our approach, as it is responsible for restoring the malware to its original form during execution. However, creating a code loader is tricky as it changes its structure depending on the execution environment. Our approach hardcodes the code loader for our experimental architecture (x86) and can be used to only attack 32-bit architectures. However, as the loaded code is just 24 bytes, it can be easily translated to other architectures like x86-64. Furthermore, the memory addresses of different sections of a file change each time it loads into memory due to the Address Space Layout Randomization (ASLR) feature. In our approach, we disabled the ASLR feature in files header characteristic flags during the injection of code cave. Once the ASLR is disabled, the file is loaded into the same base address each time in the memory. These challenges can limit the practicality of the approach.
    
    \item \textbf{Functionality Verification:} 
    Since our approach involves modifying large chunks of byte values, the chances of breaking the file are very high. To verify the functionality, we used two approaches: Debuggers to check all the debugging events and Windows virtual box to execute the adversarial malware. All the debugging events are monitored while the malware is executed inside an isolated Windows virtual box. Despite our attempt to verify a file's original behavior, we could only test if the malware executes appropriately without any errors and exceptions. It is practically impossible to capture the entire behavior of malware and verify each behavior. Due to the lack of techniques to quantify all malware behaviors before and after the AE attack, we could not check malware health in a fault-proof manner. 
\end{itemize}

\section{Conclusion}

\label{Sec:Conclusion}
In this work, we researched the vulnerability in the structure of Windows PE malware and proposed a novel way to exploit the structural weakness to carry out adversarial malware evasion attacks. We first trained MalConv and MalConv2 malware detectors to use them as our target model. We then chose a target section among the \texttt{.text}, \texttt{.data}, and \texttt{.rdata} sections of the malware PE file to inject an \textit{intra-section} code cave and optimise the generated adversarial perturbations in the inserted code caves. Additionally, we introduce the injection of a code loader to restore the original behavior of a malware file. The injected code loader that runs at the start of the malware file overwrites the perturbations and restores the original malware. We verified the functionality of an adversarial malware file by executing and debugging it. Our experiments showed promising results, outperforming existing works significantly. The results showed a maximum evasion rate of 92.31\% against MalConv and 97.93\% against MalConv2 on adding maximum perturbations of 15\% of the size of a malware file generated by using gradient descent. We further attacked the MalConv model using the fast gradient sign method to get a maximum evasion rate of 96.26\%. All our results show the significance of injecting perturbation in different sections of malware. Our work concludes by discussing the challenges and the limitations faced by our work. Clearly, the proposed intra-section code cave injection improves the performance of adversarial evasion attacks, provides flexibility in attack location, and hides the adversarial perturbation in less obvious regions of a malware file. For future work, more AE attacks on PE malware can be carried out against malware detectors other than MalConv and MalConv2, considering blackbox environments while exploring other vulnerabilities in the structure of PE files.

\section*{Acknowledgment}

\bibliographystyle{IEEEtran}
\bibliography{main}

\end{document}